\documentclass[preprint]{aastex}
\usepackage{subfig}
\usepackage[dvips]{color}

\slugcomment{Draft 20101229}
\shorttitle{Biconical Cavity in G10.6-0.4}
\shortauthors{}

\begin{document}

\title{An Overall Picture of the Gas Flow In Massive Cluster Forming Region: The Case of G10.6-0.4}

\author{Hauyu Baobab Liu\altaffilmark{1,2,3}}
\affil{Harvard-Smithsonian Center for Astrophysics, 60 Garden Street, Cambridge, MA 02138}
\email{hlu@cfa.havard.edu}

\author{Qizhou Zhang\altaffilmark{2}}
\affil{Harvard-Smithsonian Center for Astrophysics, 60 Garden Street, Cambridge, MA 02138}
\email{qzhang@cfa.harvard.edu}

\author{Paul T. P. Ho\altaffilmark{1,3}}
\affil{Academia Sinica Institute of Astronomy and Astrophysics, \\P.O. Box 23-141, Taipei, 106 Taiwan}
\email{pho@asiaa.sinica.edu.tw}



\altaffiltext{1}{Academia Sinica Institute of Astronomy and Astrophysics}
\altaffiltext{2}{Harvard-Smithsonian Center for Astrophysics}
\altaffiltext{3}{Department of Physics, National Taiwan University}

\begin{abstract}
The massive clump G10.6-0.4 is an OB cluster forming region, in which multiple UC H\textsc{ii} regions have been identified. 
In the present study, we report arcsecond resolution observations of the CS (1--0) transition, the NH$_{3}$ (3,3) main hyperfine inversion transition, the CH$_{3}$OH J=5 transitions, and the centimeter free--free continuum emissions in this region.
The comparisons of the molecular line emissions with the free--free continuum emissions reveal a 0.5 pc scale massive molecular envelope which is being partially dispersed by the dynamically--expanding bipolar ionized cavity. 
The massive envelope is rotationally flattened and has an enhanced molecular density in the mid--plane. 
In the center of this massive clump lies a  compact ($<$0.1 pc) hot ($\gtrsim$100 K) toroid, in which a cluster of O--type stars has formed.

This overall geometry is analogous to the standard core collapse picture in the low--mass star forming region, with a central (proto--)stellar object, a disk, an envelope, and a bipolar outflow and outflow cavity.
However, G10.6-0.4 has a much larger physical size scale ($\le$0.1 pc for typical low--mass star forming core).
Based on the observations, we propose a schematic picture of the OB cluster forming region, which incorporates the various physical mechanisms. 
This model will be tested with the observations of other embedded OB clusters, and with numerical simulations.

\vspace{1cm}
\end{abstract}


\clearpage
\section{Introduction }
\label{chap_introduction}
A M$\gtrsim$10 M$_{\odot}$ massive star has a short Kelvin--Helmholtz (cooling) timescale, and starts the nuclear burning while it is still deeply embedded in the dense molecular clump. 
During this earliest evolutionary stage, the massive stars can continue to increase their stellar mass via accretion of more dense molecular gas (Kahn 1974; Garay and Lizano 1999). 
Meanwhile, these massive stars produce large amounts of strong ionizing photons, and create the \textbf{Ultracompact (UC)} \textbf{H}\textsc{ii} \textbf{Regions}. 
The UV radiation (Krumholz, Mckee, Klein 2005; Peters et al. 2010), the pressure of the ionized gas (Keto 2002b; Krumholz et al. 2009; Peters et al. 2010) , and the strong stellar wind directly interact with the surrounding molecular gas.
In addition, as massive stars typically form in clusters, their environments may be also influenced by their interactions both radiatively and dynamically. 
These interactions potentially can disperse the accretion flow, unless the system is confined or trapped by the high density molecular gas with the right physical profiles (example solutions see Keto 2002b, 2003, 2007; also see Galv{\'a}n-Madrid et al. 2008 for an observational evidence of accretion). 
This is the greatest contrast to the formation of the low--mass stars, where accretion only occurs before the pre--main sequence phase when the surrounding gas is neutral.
To understand how massive stars finally attain a high stellar mass (e.g. $>$20 M$_{\odot}$), it is necessary to understand the accretion after the formation of an UC \textsc{Hii}.
\textit{One key factor in determining the relative importance between these feedback mechanisms as well as gravity, and hence the subsequent evolution of the dense molecular core/envelope, is the overall geometry/morphology of the entire system.}
For example, the density distribution of the molecular gas around massive stars determines how the molecular structures are self--shielded from the ionizing photons; and the overall geometry of the system determines how the radiation and the ionized photons can leak out of the system.

Owing to their typical large distances (a few kpc), the molecular structures around the UC H\textsc{ii} regions, are not well resolved in previous observations. 
How the accretion flows around individual stars and in the entire cluster is unknown. 
Hence how the OB stars gain mass via accretion after nuclear burning begins, is also unknown.
We perform new interferometric observations toward a well--studied O--type cluster forming region G10.6--0.4, in order to resolve the molecular structures.
The high column density and high brightness temperature of this source allow us to follow the detailed structures of the molecular gas over an extended region, at arcsecond resolutions.
Here we emphasize that the subject under consideration is the molecular and the ionized gas flow around a cluster of OB stars, rather than the accretion around a single massive (proto--)stellar object.
This is due to the still insufficient spatial resolution. 

The UC H\textsc{ii} region complex G10.6-0.4 is at a distance of 6 kpc (Caswell et al. 1975; Downes et al. 1980), and is still deeply embedded in a dense molecular core/envelope.
The brightest UC H\textsc{ii} region is extremely luminous (integrated flux is 2.6 Jy at 23 GHz) (Ho \& Haschick 1981; Sollins et al. 2005; Sollins \& Ho 2005), and contains an OB cluster within a 0.05 pc region.
Detections of water and OH masers (Genzel \& Downes 1977; Ho \& Haschick 1981; Ho et al. 1983; Hofner \& Churchwell 1996; Fish et al. 2005 ) indicate that the formation of massive stars is still ongoing.
The early NH$_{3}$, CS and C$^{18}$O line observations (Ho \& Haschick 1986; Keto, Ho, \& Haschick 1987; Keto, Ho, \& Haschick 1988; Omodaka et al. 1992; Ho, Terebey, \& Turner 1994 ) unveiled a 0.5 pc scale molecular envelope. 
These observations also suggested that the dynamics of the majority of the molecular gas have not yet been strongly disturbed by the YSOs or stellar activities.
At the 0.5 pc scale, the general motion of the molecular gas appears to be dominated by gravity, showing rotation along a flattened geometry (Liu et al. 2010).  
The inner part at the 0.1 pc scale appears to rotate faster while infalling towards the UC H\textsc{ii} region (Keto, Ho, \& Haschick 1987; Keto, Ho, \& Haschick 1988; Liu et al. 2010; Beltr{\'a}n et al. 2011).
Recombination line studies further suggest that the accretion flow continues across the H\textsc{ii} boundary, while evidence of outflow can be seen in the motions and structures of the H\textsc{ii} region at the sub--arcsecond angular scale (Keto 2002; Keto \& Wood 2006).  
Structures found in projection against the UC H\textsc{ii} continuum emission from NH$_{3}$ optical depth studies, suggest accretion via a very clumpy and geometrically thick rotating core at the 0.1 pc scale (Sollins \& Ho 2005).  

In this paper, we present the new arcsecond resolution (1$''$ corresponds to 0.03 pc at the distance of G10.6-0.4) molecular line (CH$_{3}$OH, NH$_{3}$, CS) data, together with very sensitive and 0$''$.5 resolution 3.6 centimeter  continuum observations.
The linear scale of the synthesized beams is smaller than the natural scale for self-gravity (Jeans length $\sim$0.1 pc for temperature $\sim$20--30 K and H$_{2}$ number density $\sim$10$^{5}$ cm$^{-3}$).
We see an unprecedented detailed morphology of the massive accretion flow which clarifies the geometrical relation between the hot core and the envelope. 
With the aid of the new results and the discussions on this target source in previous publications, a more thorough understanding is achieved. 

In massive star forming regions, the NH$_{3}$ inversion transitions are usually regarded as reliable molecular tracers for 20--100 K gas.
The main hyperfine components of these NH$_{3}$ lines are optically thick and may mask the detailed density structures.
The satellite hyperfine components are more optically thin (see Ho \& Townes 1983 for a review).
However, in most cases, the satellite hyperfine components of the NH$_{3}$ (J, K) = (1,1) inversion transitions are blended with the main hyperfine component; the satellite hyperfine components of the higher (J,K) level transitions are much weaker and are difficult to be observed with high resolutions.
With a factor of $\sim$100 higher Einstein A coefficient, the selected CH$_{3}$OH transitions potentially allow us to probe the denser regions.
We simultaneously observed multiple CH$_{3}$OH transitions which have comparable Einstein A--coefficients (and therefore critical densities), but trace a broad range of upper--level--energy (E$_{up}$).
These observations allow us to distinguish the hottest molecular gas directly associated with the UC H\textsc{ii} region from the cooler envelope.

The low E$_{up}$ and the modest critical density of the CS (1-0) transition provide sufficient optical depth, which allows us to trace the molecular gas to a much more extended region. 
However, in the case of G10.6-0.4, significant local structures can still be identified without being severely affected by the sidelobes of the dirty beam and the missing flux.
The high optical depth of CS (1-0) also provides better signal--to--noise ratio to the geometrically thinner expansional signatures, and provides information for the on--going feedback processes. 

The observational results are presented in Section \ref{chap_result}.
In Section \ref{chap_discussion}, we discuss the physical implication of the overall geometrical picture of the region. 
In addition, we estimate the feedback from various physical processes (radiative pressure, ionized gas pressure, stellar wind) and compare with our observational measurements.
A brief summary of the results are provided in Section \ref{chap_summary}.

\section{Observations and Data Reduction} 
\label{chap_obs}
\subsection{The Molecular Line Observations}
We observed the CH$_{3}$OH J=5 transitions in the 230 GHz band using the Submillimeter Array (SMA)\footnote{The Submillimeter Array is a joint project between the Smithsonian Astrophysical Observatory and the Academia Sinica Institute of Astronomy and Astrophysics, and is funded by the Smithsonian Institution and the Academia Sinica. }; we observed the NH$_{3}$ (3,3) hyperfine inversion transitions in the K-band using the NRAO\footnote{The National Radio Astronomy Observatory is a facility of the National Science Foundation operated under cooperative agreement by Associated Universities, Inc.} (Expanded) Very Large Array (VLA/EVLA) in the C--configuration; and we observed the CS (1-0) transition using the VLA/EVLA in the DnC--configuration.
For more on the SMA and its specifications, see Ho, Moran \& Lo (2004).
The size of the synthesized beams in these observations are comparable.
Continuum emissions are averaged from the line-free channels and then subtracted from the line data.
The properties of the selected molecular transitions are listed in Table 1;  the instrumental parameters are summarized in Table  2;  the observational settings are provided in Table 3.
The CH$_{3}$OH 5(-2,4)--4(-2,3) E and CH$_{3}$OH 5 (2,3)--4(2,2) E lines are closely blended.
The frequency separation of these two lines is only marginally resolved. 
Considering that these two lines trace similar physical conditions, and the line-widths are broad in the observed region, we regard the blended lines as one single line in this research, and denote it henceforth as CH$_{3}$OH 5(2,3)--4(2,2).

Basic calibrations are carried out in the \texttt{MIR} package for the SMA data; and are carried out in the \texttt{AIPS} package for the the VLA/EVLA data.
All data are additionally self--calibrated in the \texttt{AIPS} package.
Imaging is also carried out in the \texttt{AIPS} package.

\subsection{The Centimeter Continuum Observations}
We observed the X band continuum emission toward G10.6-0.4 in the VLA A--configuration including the VLBA Pie--Town antenna on 2005 January 2, and in the VLA/EVLA C--configuration on 2009 July 27.
 We observed 1331+305, 0319+415, and 1820-254 as absolute flux, passband, and gain calibrators.
 
The basic calibrations, self-calibration, and imaging of these data are carried out in the \texttt{AIPS} package.
We combine the A--array+Pie--Town visibility data with the C-array visibility data, yielding a 0$''$.5$\times$0$''$.4 synthesized beam with a position angle of -43.2$^{o}$.
The observed RMS noise of the 3.6 cm continuum image is about 0.24 mJy/beam ($\sim$21 K in brightness temperature), and the peak brightness temperature is about 8700 K.
We note that the 3.6 cm continuum emission is free--free emission based on spectral index measurements. 
Hence its high brightness temperature indicates that the emission is moderately optically thick.

\section{Results}
\label{chap_result}
The observing results are separately presented in two sections.
In Section \ref{chap_geometry}, we focus on discussions of the warm and dense molecular gas, mainly traced by the NH$_{3}$ and the CH$_{3}$OH emissions. 
In Section  \ref{chap_dynamics}, we examine the dynamics of the more extended components, based on the observations of the CS (1--0) transition. 
Cross comparisons of the molecular lines have been omitted, when there are no additional significant physical implications. 

\subsection{The NH$_{3}$ and the CH$_{3}$OH Emissions}
\label{chap_geometry}
\subsubsection{The Distribution and the Dynamics of the Dense Gas}
\begin{figure}
\rotatebox{-90}{
\includegraphics[scale=0.75]{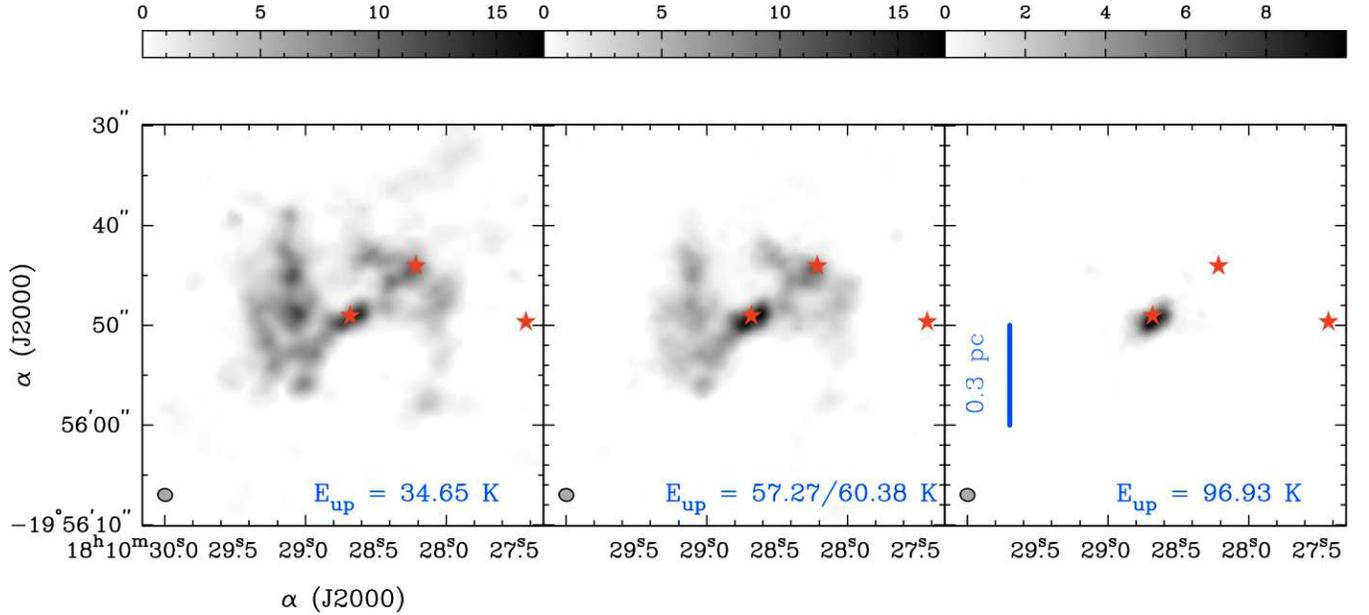}}
\caption{The velocity integrated maps of the CH$_{3}$OH J=5 transitions (\textbf{Left:} CH$_{3}$OH 5(0,5)--4(0,4) A+; \textbf{Middle:} the blended CH$_{3}$OH 5(-2,4)--4(-2,3) E and CH$_{3}$OH 5 (2,3)--4(2,2) E; \textbf{Right:} CH$_{3}$OH 5(-3,3)--4(-3,2) E).
Color--bars have the unit of Jy/beam$\cdot$kms$^{-1}$.
Three 1.3 cm free-free continuum peaks (R.A.=18$^{h}$10$^{m}$28$^{s}$.683 Decl=-19$^{o}$55$'$49$''$.07; R.A.=18$^{h}$10$^{m}$28$^{s}$.215 Decl=-19$^{o}$55$'$44$''$.07 ; R.A.=18$^{h}$10$^{m}$27$^{s}$.435 Decl=-19$^{o}$55$'$44$''$.67) are marked by red stars. 
The synthesized beams are shown in the bottom left corner of the panels.
}
\label{fig_mnt0}
\end{figure}

\begin{figure}
\includegraphics[scale=0.85]{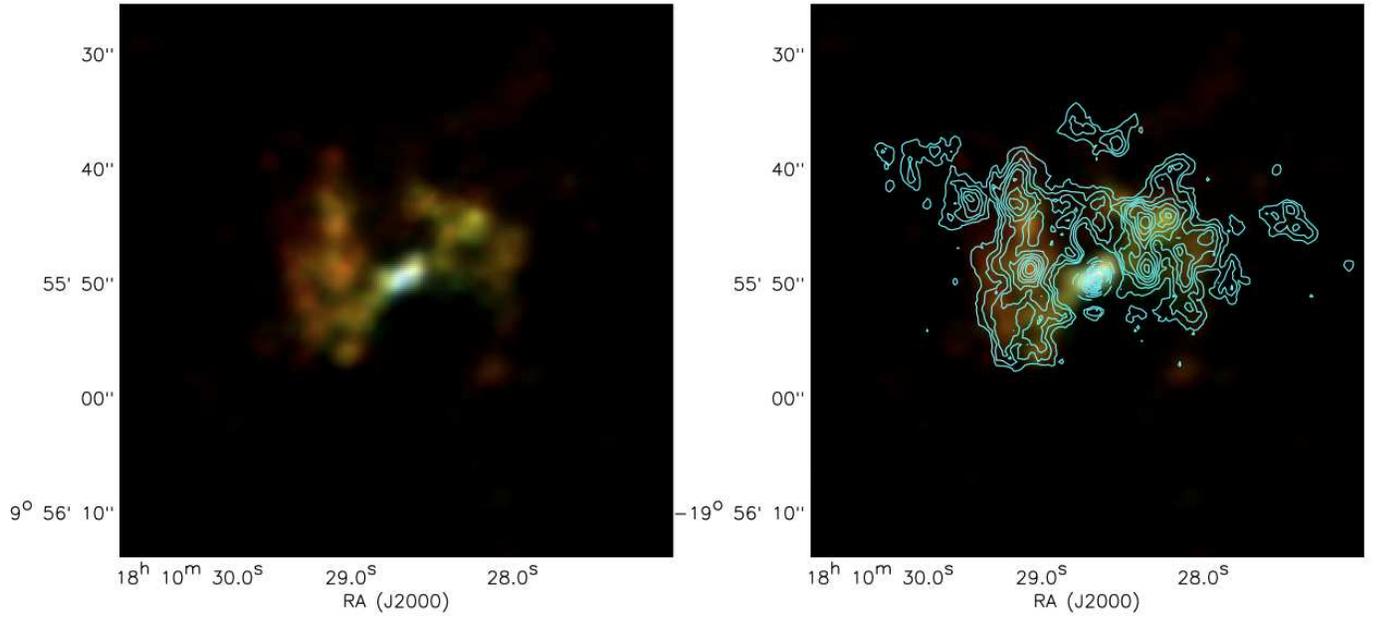}
\caption{
\textbf{Left: } The RGB image of the CH$_{3}$OH J=5 transitions (\textbf{R:} CH$_{3}$OH 5(0,5)--4(0,4) A+; \textbf{G:} the blended CH$_{3}$OH 5(-2,4)--4(-2,3) E and CH$_{3}$OH 5 (2,3)--4(2,2) E; \textbf{B:} CH$_{3}$OH 5(-3,3)--4(-3,2) E).
\textbf{Right: } The RGB image of the CH$_{3}$OH J=5 transitions overlaid with the NH$_{3}$ (3,3) main hyperfine inversion emission (contour). Solid contours start from 10\% of the emission peak with 10\% intervals; dashed contours start from -200\% of the emission peak with -200\% intervals.
Note the primary beams of the SMA ($\sim$1$'$) and the VLA ($\sim$2$'$)  observations are much larger than the angular size scale of the detected structures.
}
\label{fig_nh3}
\end{figure}

Figure \ref{fig_mnt0} shows the velocity integrated maps of the CH$_{3}$OH 5(0,5)--4(0,4) A+ transition, the blended CH$_{3}$OH 5(-2,4)--4(-2,3) E and CH$_{3}$OH 5(2,3)--4(2,2) E transitions, and the CH$_{3}$OH 5(-3,3)--4(-3,2) E transition.
In Figure \ref{fig_nh3}, we present these three images in RGB colors, and an overlay with the velocity integrated map of the NH$_{3}$ (3,3) main hyperfine line.
In the projected area of the UC H\textsc{ii} region, the NH$_{3}$ (3,3) main hyperfine line is detected in strong absorption (see Keto, Ho \& Haschick 1987; Keto, Ho, \& Haschick 1988; Sollins et al. 2005; Liu et al. 2010 for more descriptions) and is presented with much larger contour intervals. 
We see that:

\begin{itemize}
\item The highest excitation CH$_{3}$OH transition (blue) exclusively traces a compact ($\sim$0.1 pc) flattened hot toroid in the center ($\sim$300K, Klaassen et al. 2009; $\sim$87 K, Beltr{\'a}n et al. 2011). 
The  lower excitation CH$_{3}$OH transitions trace the warm ($\sim$50 K; Keto, Ho \& Haschick 1987) and dense molecular gas to an extended ($\sim$0.5 pc) region. 
The ratio of the velocity integrated flux of the  CH$_{3}$OH 5(0,5)--4(0,4) A+ transition to that of the CH$_{3}$OH 5(-3,3)--4(-3,2) E transitions in the hot toroid region is about 0.56. 
In the optically thin limit, this line ratio reflects an averaged excitation temperature of $\sim$500 K.
If the lower excitation line is saturated owing to high optical depth, the excitation temperature would be overestimated.
\item  We see the complicated structures from the extended CH$_{3}$OH emissions.
 The majority of the CH$_{3}$OH emission has a southeast--northwest distribution. 
 Two low column density cavities are clearly seen northeast and southwest of the hot toroid.
 \end{itemize}

We note that the size scale of the region where we detect significant emissions is small as compared with the angular size scale of the SMA primary beam ($\sim$1$'$)  and that of the VLA primary beam ($\sim$2$'$).
Since the detected structures are well centered within the primary beams, they are not affected by the primary beam attenuation.
We therefore do not perform primary beam corrections to the maps. 
The shortest baselines of the CH$_{3}$OH and the NH$_{3}$ data sets (Table \ref{table_molecule_list2}) correspond to the maximum angular size scale of $\sim$35$''$ and $\sim$90$''$, respectively.
Hence the absence of structures with angular size scales $>$20$''$ suggests that the compact appearance of the emissions is physical in nature, and is  not caused by sampling defects.

\begin{figure}
\rotatebox{-90}{
\includegraphics[scale=0.75]{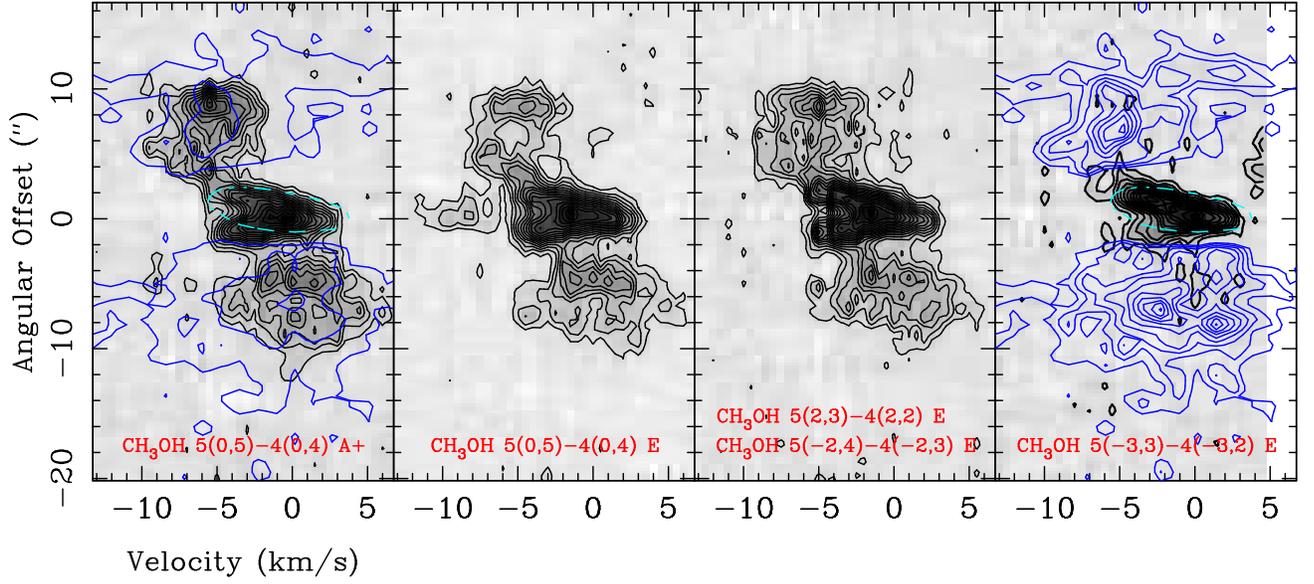}}
\caption{The PV diagrams of the CH$_{3}$OH J=5 transitions (grey scale and black contours) cut in the plane of rotation, which is centered at the coordinates of R.A. = 18$^{h}$10$^{m}$28.64$^{s}$ and Decl = -19$^{o}$55$'$49.22$''$ with position angle pa = 140$^{o}$
From left to right is the CH$_{3}$OH 5(0,5)--4(0,4) A+ transition, the CH$_{3}$OH 5(0,5)--4(0,4) E transition, the blended CH$_{3}$OH 5(-2,4)--4(-2,3) E and CH$_{3}$OH 5 (2,3)--4(2,2) E transitions, and the CH$_{3}$OH 5(-3,3)--4(-3,2) E. The black contour intervals and the first contour level are 0.18 Jy/beam (1.9 K) for each panel.  In the right most panel, we also plot the NH$_{3}$ (3,3) main hyperfine emission line PV diagram in blue contours, start from 0.01 Jy/beam (11 K) with 0.01 Jy/beam intervals; and plot the NH$_{3}$ (3,3) main hyperfine absorption line by one dashed green contour at the level of -0.5 Jy/beam. 
We plot the 0.01, 0.04, and -0.5 Jy/beam contour levels of the NH$_{3}$  (3,3) main hyperfine line in the left most panel for comparison. 
}
\label{fig_ch3ohpv}
\end{figure}

From  Figure \ref{fig_nh3}, we see that most structures traced by the CH$_{3}$OH emissions have their NH$_{3}$ counterparts; however, the NH$_{3}$ emission shows more significantly detected structures than the CH$_{3}$OH emissions.
A similar effect appears in the velocity domain. 
Figure \ref{fig_ch3ohpv} shows the position-velocity (PV) diagrams of the NH$_{3}$ (3,3) main hyperfine line and all selected CH$_{3}$OH J=5 transitions.
The PV cut is centered on the brightest free-free continuum source\footnote{This is exactly the same PV cut which shows the maximum velocity gradient (Liu et al. 2010).} at the coordinates of R.A. = 18$^{h}$10$^{m}$28.64$^{s}$ and Decl = -19$^{o}$55$'$49.22$''$ with a position angle PA = 140$^{o}$ (the positive angle is defined in the usual counterclockwise fashion).
From these PV diagrams, we see that:
\begin{itemize}
\item The CH$_{3}$OH 5(-3,3)--4(-3,2) E transition has the highest upper--level energy E$_{up}$ of 96.9 K. The PV diagram of this transition reveals the fast rotating hot toroid. 
\item The CH$_{3}$OH transitions with lower upper--level  energies trace the molecular gas in a more extended region. A significant velocity gradient is also seen in the extended emission. 
\item From the left most panel of Figure \ref{fig_ch3ohpv}, we see that the dominant emission of the NH$_{3}$ (3,3) main hyperfine inversion line shows a good agreement with the CH$_{3}$OH emissions. 
The NH$_{3}$ (3,3) absorption line consistently traces the fast rotating hot toroid as seen in the PV diagram of the CH$_{3}$OH 5(-3,3)--4(-3,2) E transition. 
Additionally, the NH$_{3}$ (3,3) emission shows significant detections of broad line emission features (e.g. at 7$''$ and 11$''$) and some diffuse broad line emission (e.g. around -3$''$-- -11$''$).
\end{itemize}

The NH$_{3}$ molecule has a stable abundance over a large range of densities and temperature without depletion onto grains. 
Hence its emission lines can be used to trace the dynamics in the bulk of the dense molecular accretion flow.
The detected velocity gradients can be explained by the marginally centrifugally supported motion of the molecular gas, in response to the gravity of the enclosed molecular and the stellar mass (see Liu et al. 2010 for more discussions).
The projected area and the rotational motions traced by the CH$_{3}$OH are highly consistent with those which are traced by the NH$_{3}$ emissions.
This suggests that these two molecules are tracing dense gas in the same region.
Nevertheless, there are differences in their detailed brightness distributions which suggest that in this particular source, the CH$_{3}$OH emissions are tracing different excitation/chemical conditions (e.g. temperature), in the same flow traced by NH$_{3}$ (see also the discussions for the CH$_{3}$OH emissions in Liu, Ho \& Zhang 2010).
In addition, the abundance of the CH$_{3}$OH molecule may also be affected by outflow. 

The NH$_{3}$ (3,3) main hyperfine inversion transition has comparable upper--level energy (E$_{up}$/k = 124.5 K) with the highest excitation CH$_{3}$OH 5(-3,3)--4(-3,2) E transition (E$_{up}$/k = 96.9 K), but in contrast it shows an extended distribution.
One explanation is that the NH$_{3}$ molecule has a higher abundance ([NH$_{3}$]/[H$_{2}$] = 10$^{-7}$; Keto, Ho \& Haschick 1987) than the CH$_{3}$OH molecule ([CH$_{3}$OH]/[H$_{2}$] = 2$\cdot$10$^{-9}$; Takakuwa, Ohashi, \& Hirano 2003), and thus becomes optically thick and shows strong emission. 
The excellent match between the NH$_{3}$ emission and the lowest excitation CH$_{3}$OH emission suggests that the emissions of both molecules trace the real physical structures. 
We know that for the dense gas with temperature around 50 K, the statistical weight compensates the Boltzmann factor, and therefore the population of NH$_{3}$ at the (3,3) level can be comparable to the population at the lowest excitation (1,1) level.
The CH$_{3}$OH emissions and the NH$_{3}$ (3,3) main hyperfine emission may both trace a population of molecular gas which is significantly warmer and denser than the ambient molecular gas.
\textit{We call this warm ($\ge$30 K) and dense ($>$10$^{5}$ cm$^{-3}$) region the massive envelope hereafter.}
The previous observations on C$^{18}$O (2--1) unveiled a parsec scale filamentary structure in the south of the massive envelope (Ho, Terebey, \& Turner 1994).
The distribution of the molecular mass on an even larger scale has to be further examined by sensitive single--dish maps.

The selected CH$_{3}$OH transitions have higher Einstein A--Coefficients than the NH$_{3}$ (3,3) main hyperfine inversion transition, and potentially only trace the denser components.
The observations in the clearest example of photon-dominated regions (PDR), the Orion Bar, further suggest that the CH$_{3}$OH emissions trace the clumpy materials, which have an H$_{2}$ volume density higher than 10$^{6}$ cm$^{-3}$, and are self-shielded from photo ionization (Leurini et al. 2010).
The abundance of the CH$_{3}$OH molecule can be greatly enhanced if the molecular gas is shocked by high velocity outflows with velocities greater than $\sim$20 kms$^{-1}$ (see discussions in Takakuwa, Ohashi, \& Hirano 2003).
We therefore cannot rule out the possibilities that zones with locally enhanced CH$_{3}$OH abundance/emission are distributed over the entire region.  

Part of those extended NH$_{3}$ emissions can also be attributed to the externally heated lower density surfaces of the extended structures.
Other possibilities will have to be investigated by higher resolution observations which define the detailed temperature and density profiles of individual local structures.

\subsubsection{Comparison with the Distribution of the Free--Free Continuum Emissions}
\begin{figure}
\includegraphics[scale=0.6]{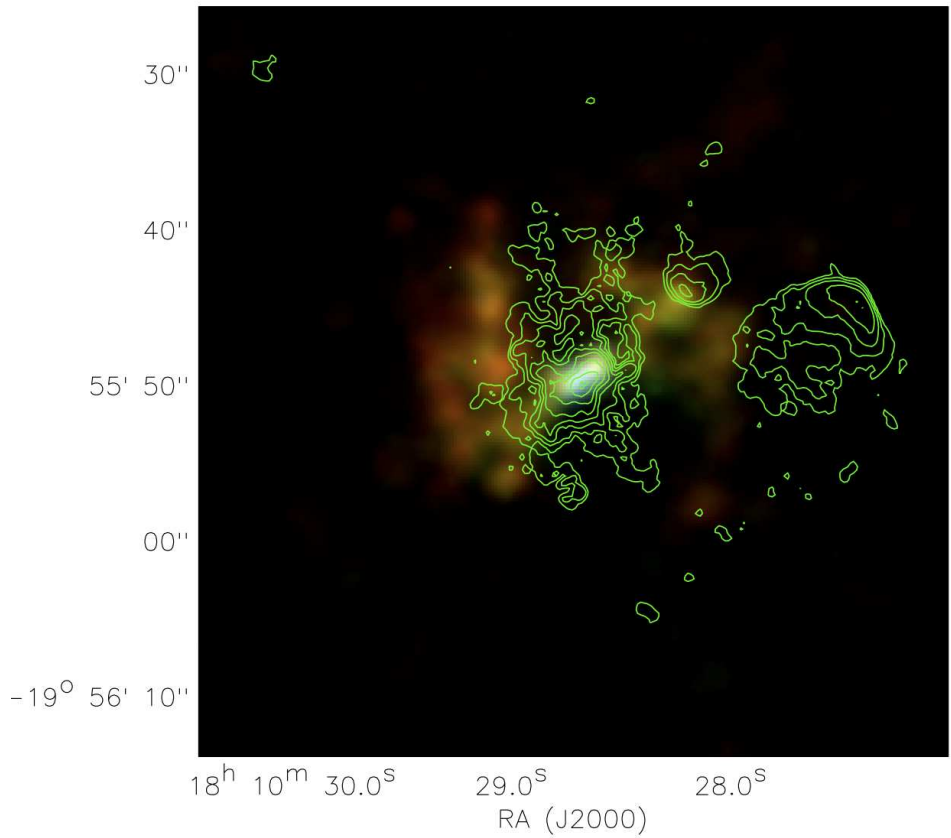}
\caption{
The RGB image of the CH$_{3}$OH J=5 transitions overlayed with the 3.6 cm continuum image (contour). Contours show the levels of [1, 2, 4, 8, 16, 32, 64, 96]\% of the peak intensity of 0.1 Jy/beam. 
}
\label{fig_ionout}
\end{figure}

In Figure \ref{fig_ionout}, we compare the distribution of the molecular gas traced by the CH$_{3}$OH transitions with the distribution of the ionized gas traced by the 3.6 cm free--free continuum emission.
By comparing the left panel of Figure \ref{fig_nh3} with this figure, we find that the 3.6 cm free-free continuum emissions peak at the hot toroid. 
The distribution of the extended free-free continuum emission fits into the projected area of the low column density cavities seen in the CH$_{3}$OH RGB map.

These observed features suggest an overall geometry of a core--envelope--cavity system.
This overall geometry resembles the standard disk--envelope system in the low mass ($\le$1 M$_{\odot}$) star forming region with bipolar outflow and outflow cavity, except on a larger and more massive scale.
This overall geometry remains even after the embedded OB cluster starts the nuclear burning, and other physical processes start to affect the dynamical evolution (Section \ref{chap_introduction}).
We emphasize that the subject under consideration is the formation and accretion of a cluster of OB star. 
By comparing the flux of the ionizing photons and the bolometric luminosity, Ho and Haschick (1981) suggested that a few massive stars (from 06.5 to B0) have already been formed and are embedded in the central 0.1 pc region (see also the discussions in Keto 2002, Sollins \& Ho 2005, and Keto \& Wood 2006).
This cluster of OB stars contains $\sim$200 M$_{\odot}$ of stellar mass in total.
As pointed out by Sollins and Ho (2005), the source G10.6-0.4 stands distinct from the cases where a single massive star dominates at the center  (e.g. IRAS 20126, IRAS 18089, G192.16, etc. Zhang et al. 1998; Beuther et al. 2004; Shepherd et al. 2001), which might have formed in a process
similar to low--mass stars (Keto \& Zhang 2010).

In G10.6-0.4, a parsec scale massive envelope ($\ge$2500 M$_{\odot}$ over the central 30$''$ region; Ho, Terebey \& Turner 1994) appears to be undergoing a coherent motion in response to the gravity of the enclosed mass and the specific angular momentum. 
In this case, the mass of the embedded OB stars is a less significant fraction ($<$10\%) as compared to the molecular mass in the envelope.
The fact that a concentration of OB stars formed in the small central 0.1 pc region suggests that the global contraction of the massive envelope may have been efficient.
We note that in another OB cluster forming region G20.08-0.14N (L$\sim$6.6$\cdot$10$^{5}$ L$_{\odot}$, d$\sim$12.3 kpc), the emissions of the NH$_{3}$ molecules and various hot core tracers also show a similar geometry and kinematic profile with those in G10.6-0.4 (Galv{\'a}n-Madrid et al. 2009).

\subsection{The CS (1--0) Emissions}
\label{chap_dynamics}
\subsubsection{Distribution}
\label{chap_csdis}
\begin{figure}
\includegraphics[scale=0.85]{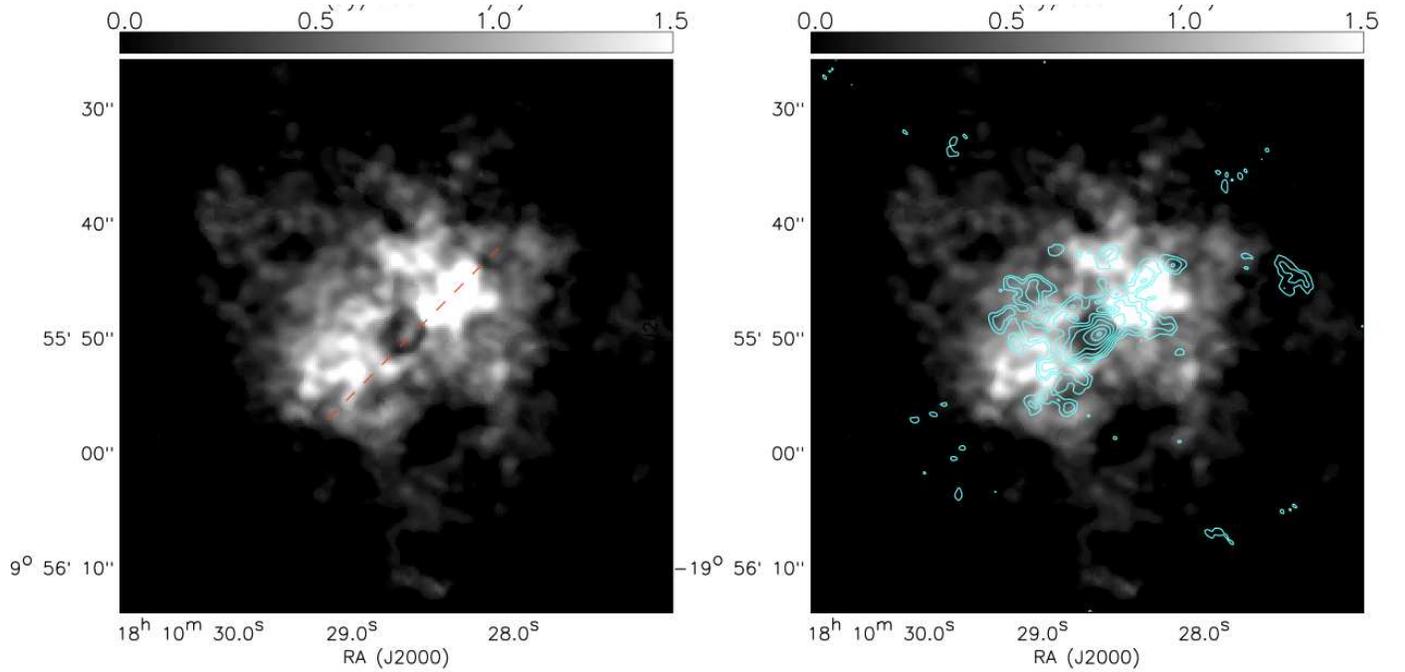}
\caption{
\textbf{Grey Scale: } The moment 0 map of the CS (1--0) emission. The synthesized beam of this map is 1.4$''$$\times$1.2$''$, with position angle 77$^{o}$.
\textbf{Contour: } The 1.3 mm continuum image. This data is taken by SMA subcompact+compact+very extended array with 0$''$.79$\times$0$''$.58 resolution. 
 Contours show the levels of [1, 2, 4, 8, 16, 32, 64, 96]\% of the peak brightness temperature 49 K. 
A proposed plane of rotation is indicated by a red dashed line. The red dashed line passes through two free-free continuum peaks, and has a position angle of 135$^{o}$, which is consistent with the position angle of the plane of rotation of the central 0.1 pc scale hot core (140$^{o}$$\pm$5$^{o}$, Liu et al. 2010).
}
\label{fig_cs}
\end{figure}

\begin{figure}
\includegraphics[scale=0.7]{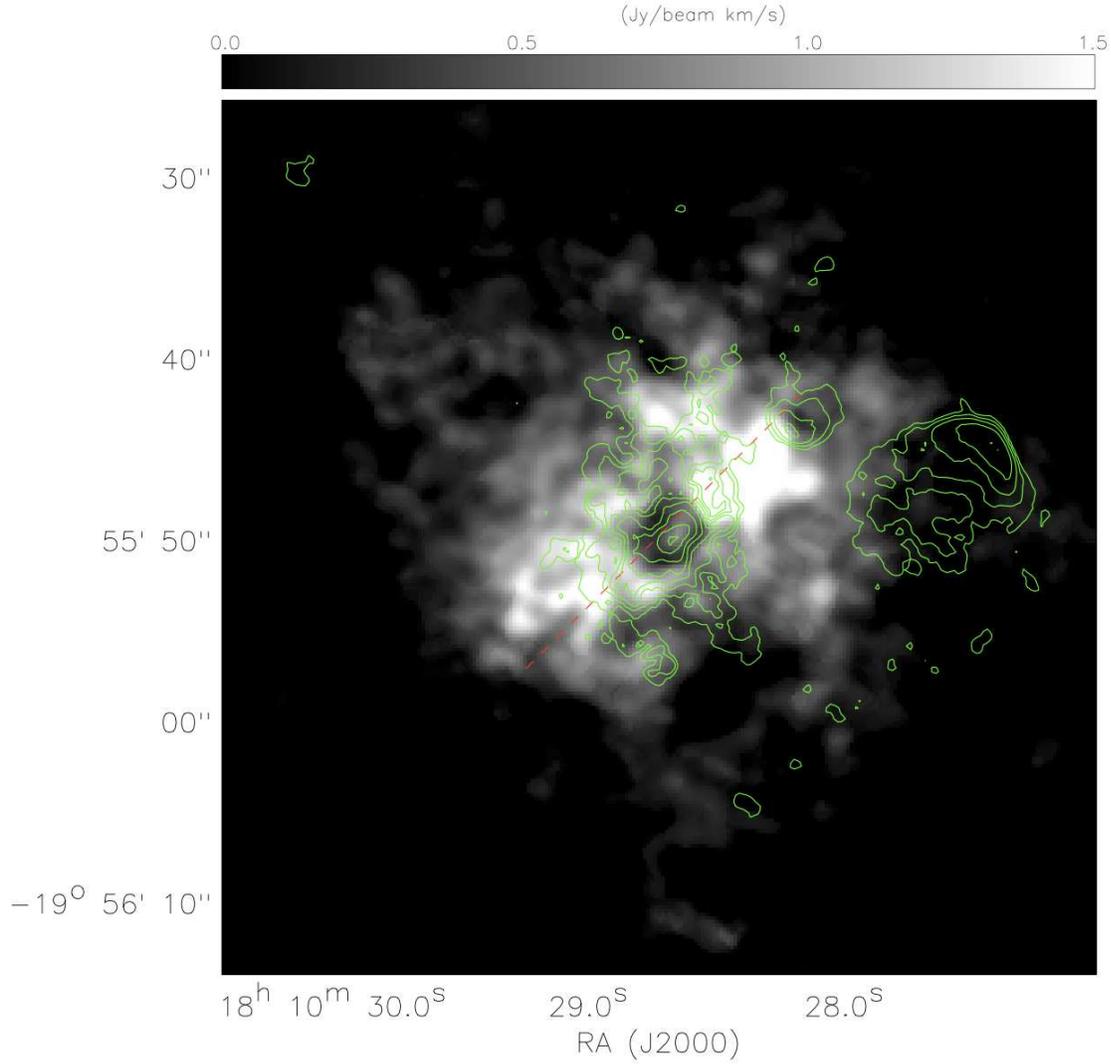}
\caption{ 
\textbf{Grey Scale: } The moment 0 map of the CS (1--0) emission. The synthesized beam of this map is 1.4$''$$\times$1.2$''$, with position angle 77$^{o}$.
\textbf{Contour: } The 3.6 cm free-free continuum image; resolution 0$''$.5$\times$0$''$.4.
 Contours show the levels of [1, 2, 4, 8, 16, 32, 64, 96]\% of the peak intensity of 0.1 Jy/beam. 
}
\label{fig_csx}
\end{figure}

Figures \ref{fig_cs} and \ref{fig_csx} show the velocity integrated (moment 0) map of the CS (1--0) emission,  the 1.3 mm continuum image, and the 3.6 cm continuum image. 
In the 3.6 cm continuum image (Figure \ref{fig_csx}), we see three isolated UC H\textsc{ii} regions (from east to west, UC H\textsc{ii} region A, B, C hereafter), of which we also detect their free-free continuum emission at 1.3 mm (Figure \ref{fig_cs}, right panel). 

The visibilities of the CS (1--0) data cover a range of uv--distances of 4--245 $k\lambda$, which correspond to $\sim$52$''$--0$''$.8 in angular scale.
The 1$\sigma$ rms in the CS velocity integrated map is $\sim$77 mJy/beam$\cdot$kms$^{-1}$ (24 K$\cdot$kms$^{-1}$).
The achieved brightness sensitivity is comparable to the theoretical rms noise (also achieved in the channel maps).
With the presence of the bright UC H\textsc{ii} regions, the detection of CS (1--0) is locally biased by the continuum subtraction.
Structures distributed in front of the UC H\textsc{ii} regions will be detected in emission line only if they emit brighter CS (1-0) emission than the local free-free continuum background; otherwise, they will be detected in absorption line. 
We omit plotting the negative flux due to absorption to enhance the contrast of emission features in the gray scale.
Structures behind the UC H\textsc{ii} region are not affected by the continuum subtraction and are always detected in emission line, unless the 49 GHz continuum emission is optically thick.
In regions where the brightness temperature of the free-free continuum emission is comparable with the excitation temperature of the molecular gas, the absolute values of the flux density of the emission line and the absorption line are comparable.
Smoothed by the finite synthesized beam, the emission line and the absorption line signatures may cancel each other, and reduce the signal--to--noise ratio.


From Figures \ref{fig_cs} and \ref{fig_csx} we see:
\begin{itemize}
\item The brightest regions of the CS (1--0) emission agrees well with the 1.3 mm continuum emission.
(The significant deficit of CS (1--0) emission in the middle of the maps is due to the subtraction of the bright free-free continuum in the UC H\textsc{ii} region.)
\item  We see an extended dense flattened structure with a size scale of $\sim$0.5 pc. 
The orientation is indicated by the red dashed line, which has a position angle of 135$^{o}$, and is consistent with the orientation of the central 0.1 pc hot toroid (Liu et al. 2010).
\item Northeastern to the dense flattened structure, the extended CS emission shows filamentary structures.
The size scale of those filamentary structures is a few times of 0.1 pc.
Southwestern to the dense flattened structure, we see an upside--down \textbf{V} shaped cavity--wall signature.
\end{itemize}

The 1.3 mm continuum emission arises predominantly from the free-free continuum emission in the central 2$''$--3$''$ UC H\textsc{ii} region; in the extended region, the 1.3 mm continuum emission traces the thermal dust emission (Liu et al. 2010; Liu, Ho \& Zhang 2010).
The thermal dust emission is usually regarded as a reliable tracer of the molecular hydrogen. 
The general agreement between the CS (1--0) image and the 1.3 mm continuum image indicates that CS (1--0) also traces the molecular hydrogen. 
However, in the optically thick case, the velocity integrated flux of CS (1--0) can be biased by saturation and/or foreground absorption, and also by the non--uniform excitation temperature.
These explain some inconsistency between the CS (1--0) velocity integrated image and the 1.3 mm continuum image.
By inspecting the position--velocity diagrams (Section \ref{chap_pv}), we do not see a gap of emission line, and therefore we argue that the self--absorption does not severely affect the overall picture of geometry. 
The excitation temperature only changes the relative strength of emissions, and should not greatly alter the overall picture either, as long as the excitation does not change by a very large factor.

The distribution of the CS (1--0) emissions is consistent with the overall geometry suggested by our observations of the CH$_{3}$OH J=5 transitions.
However, instead of seeing a clean biconical feature as in the CH$_{3}$OH maps, we detect significant CS (1-0) emission in the projected area of the biconical cavity.
This can be due to the fact that the CS (1--0) transition has a much lower upper--level--energy (E$_{up}$ = 2.35 K), and a factor of $\sim$50 lower Einstein A--Coefficient. 
These two factors make CS easily excited in the outer layer of the cavity--wall with lower temperature (and molecular gas density). 
This property allows us to robustly detect the dynamics of the cavity--wall, and explicitly measure the expansional velocity.

\begin{figure}
\rotatebox{-90}{
\includegraphics[scale=0.53]{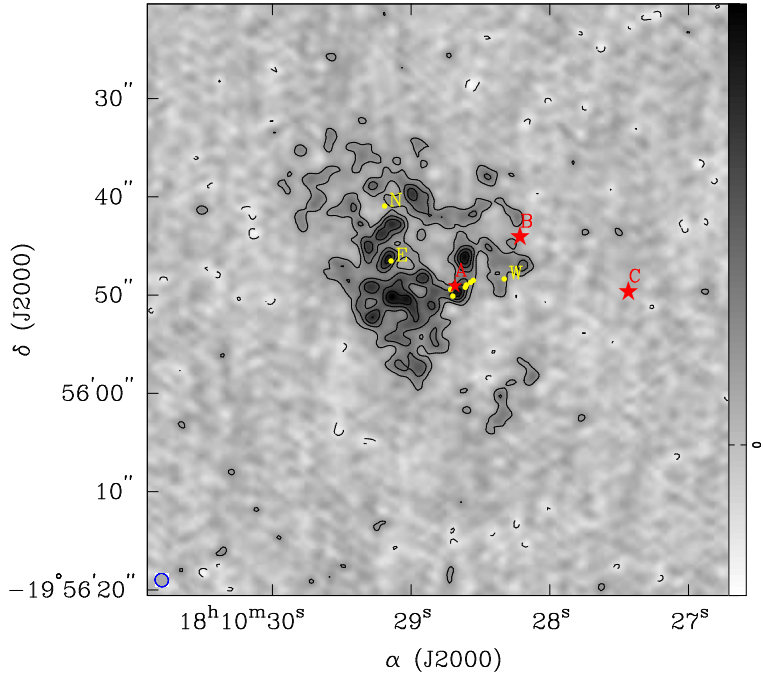}}
\caption{The example channel map at v$_{lsr}$= -7.8 kms$^{-1}$. Contours start from -0.06 Jy/beam with 0.06 Jy/beam intervals. The synthesized beam is shown in the bottom left corner of this image.
Three 1.3 cm free-free continuum peaks (R.A.=18$^{h}$10$^{m}$28$^{s}$.683 Decl=-19$^{o}$55$'$49$''$.07; R.A.=18$^{h}$10$^{m}$28$^{s}$.215 Decl=-19$^{o}$55$'$44$''$.07 ; R.A.=18$^{h}$10$^{m}$27$^{s}$.435 Decl=-19$^{o}$55$'$44$''$.67) are marked by red stars. 
Dot symbols mark the water maser detections (Hofner and Churchwell 1996). Three water maser sources which are not spatially associated with the UC H\textsc{ii} region A are labeled by N, W, and E, respectively.
The relative positional accuracy of the maser data is about 0.1$''$, which is smaller than the size of the Dots.
}
\label{fig_chm78}
\end{figure}

Our CS (1--0) maps additionally unveil an unprecedentedly detailed morphology in the massive cluster forming clump.
The velocity integrated map apparently shows abundant structures with a size scale of a few arcseconds (0.03 pc). 
In figures \ref{fig_chm78}, we present an example channel image at $v$$_{lsr}$= -7.8 kms$^{-1}$.
In this channel, we do not detect compact (2$''$--3$''$) absorption features towards the brightest UC H\textsc{ii} region A, which alleviates the potential confusion from the sidelobes of a strong source. 
We see that the arcsecond scale structures have a high contrast with respect to the background diffuse emission.
We note that it is unlikely that our images can achieve the theoretical rms noise level if the extended diffuse emission has dominant contributions to the total flux.

\begin{figure}
\rotatebox{-90}{
\includegraphics[scale=0.7]{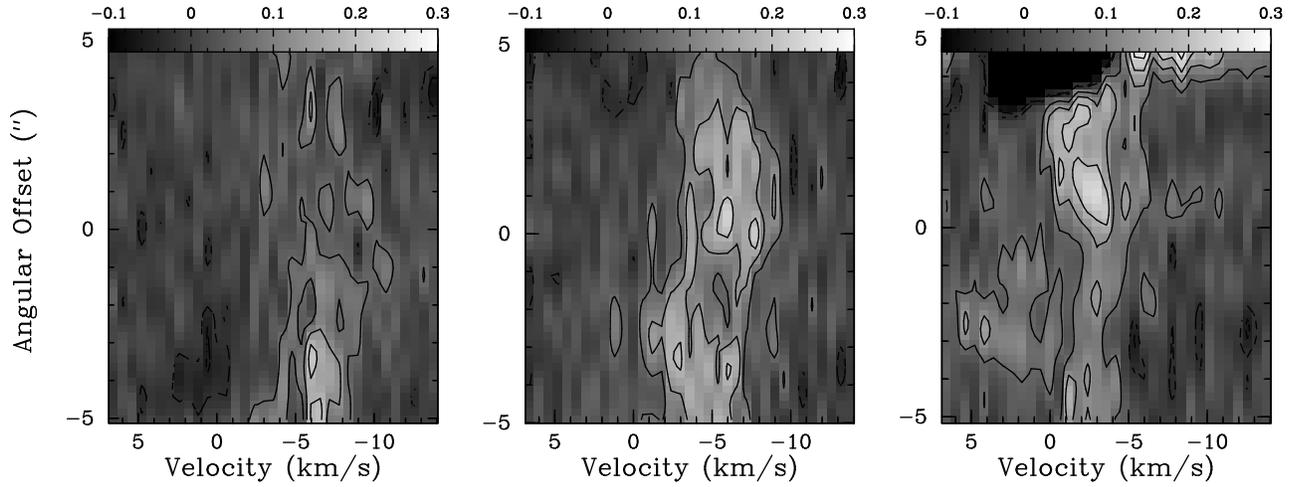}}
\caption{
From left to right, the position-velocity (PV) diagrams at the locations of water maser sources N, E, and W.
Contours start from -0.06 Jy/beam with 0.06 Jy/beam intervals. The synthesized beam is shown in the bottom left corner of this image.
Positive angular offset is defined in the east.
The PV cuts center at the locations of the water maser sources, and have the position angle PA=90$^{o}$.
Color--bars have the unit of Jy/beam.
The deep dark region in the right most panel is the absorption line against the bright background continuum emission.
}
\label{fig_pvwater}
\end{figure}

The small scale emissions may come from compact dense molecular cores, or from local structures with high temperature or high CS abundance owing to the interactions with the (proto--)stellar activities and shocks.
The compact dense molecular cores are potentially the future or current sites of star formation.
In Figure \ref{fig_chm78}, we see that two significant local maxima of CS (1--0) emission are spatially associated with water maser sources (water maser E, and the cluster of water maser sources around the UC H\textsc{ii} region A; Hofner \& Churchwell 1996).
Figure \ref{fig_pvwater} shows the PV diagrams of three cuts through the water maser sources N, E, and W.
From this figure we see detections of bright components near ($\le$1$''$) the water maser sources E and W.
The water maser source E may be associated with the local maxima of CS (1--0) emission at -7.8 kms$^{-1}$, or -5.4 kms$^{-1}$.
We do not find bright local CS (1--0) emission peak around the water maser source N, which may be due to the fact that its parent molecular core is already dispersed by the outflows.
Recalling the small high opacity clumps detected in the NH$_{3}$ (3,3) hyperfine inversion line absorption (Sollins and Ho 2005), it is not surprising that more of them are detected in emission line.
However, our current data are insufficient to robustly distinguish the dense cores from the local shock and outflow signatures, which leads to the difficulties in a quantitative analysis of the core mass function. 
In the other paper on high velocity molecular outflows in this region, we will discuss the feedback from those local sites of star formation and the implications of clustering of stars (Liu, Ho \& Zhang 2010).

\subsubsection{Dynamical Motions}
\label{chap_pv}
\begin{figure}
\rotatebox{-90}{
\includegraphics[scale=0.7]{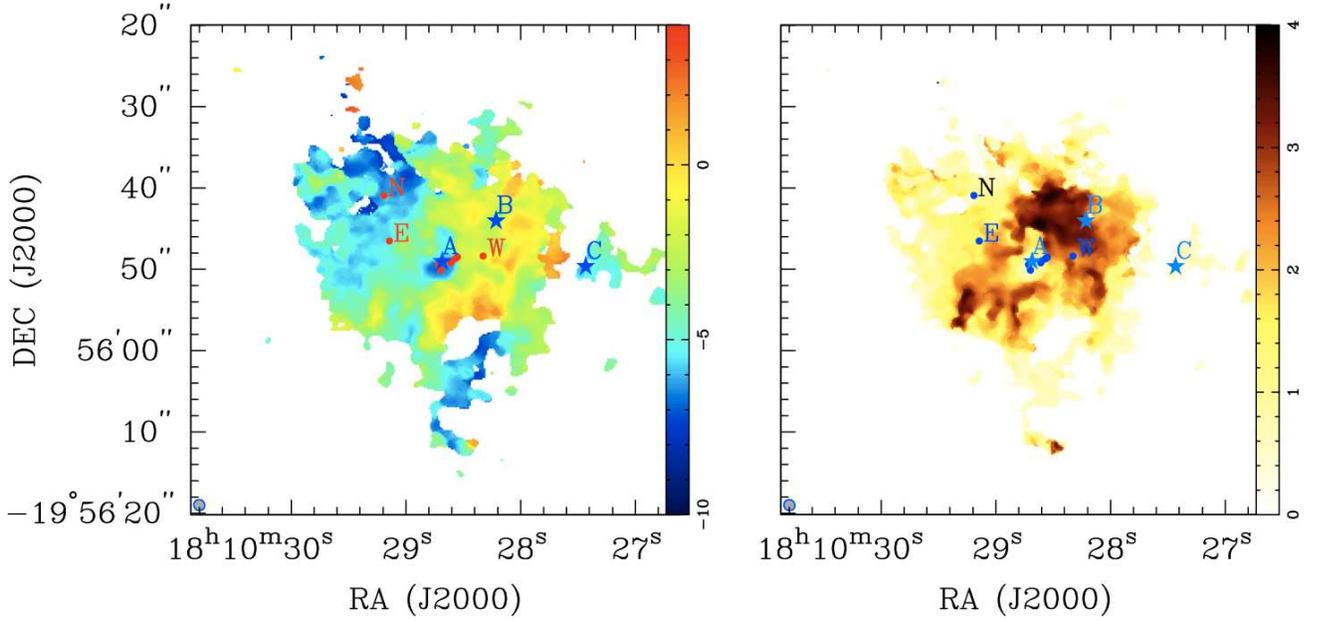}}
\caption{
The mean velocity map (left) and the velocity dispersion map (right) of the CS (1--0) emission.
Color--bars have the unit of kms$^{-1}$.
Three 1.3 cm free-free continuum peaks (R.A.=18$^{h}$10$^{m}$28$^{s}$.683 Decl=-19$^{o}$55$'$49$''$.07; R.A.=18$^{h}$10$^{m}$28$^{s}$.215 Decl=-19$^{o}$55$'$44$''$.07 ; R.A.=18$^{h}$10$^{m}$27$^{s}$.435 Decl=-19$^{o}$55$'$44$''$.67) are marked by blue stars. 
Dot symbols mark the water maser detections (Hofner and Churchwell 1996). Three water maser sources which are not spatially associated with the UC H\textsc{ii} region A are labeled by N, W, and E, respectively.
The relative positional accuracy of the maser data is about 0.1$''$, which is smaller than the size of the Dots.
}
\label{fig_mnt12}
\end{figure}

We show the mean velocity map (moment one) and the velocity dispersion map (moment two) of the CS (1--0) emission in Figure \ref{fig_mnt12}.
From Figure \ref{fig_mnt12}, we see that the entire system has a bimodal velocity profile. 
In addition to the general rotation with velocity gradient from the southeast to the northwest, we see blueshifted gas in the northeast and redshifted gas in the southwest.
These additional blueshifted and redshifted gas cover a larger velocity range than the general rotational motion, and the orientation resembles a bipolar expansion with a certain inclination.  
Geometrically, the upside--down \textbf{V} shaped cavity--wall signature in the southwest can also be explained by the bipolar expansion. 
We note that our 3$\sigma$ detection limit is about 18 K, which is only sensitive to structures which are sufficiently warm and have high optical depth.
The moment one and moment two maps thus can be biased due to the cutoff of the fainter high velocity line wings, and over interpretations should be omitted. 

To understand the dynamics of the cavity--wall, we make two PV cuts through the southern part of the biconical cavity (centered at \textbf{cut1: } R.A. = 18$^{h}$10$^{m}$28.51$^{s}$ , Decl = -19$^{o}$55$'$54.3$''$, and \textbf{cut2: }  R.A. = 18$^{h}$10$^{m}$28.48$^{s}$ , Decl = -19$^{o}$56$'$1.0$''$) with a position angle 140$^{o}$ (consistent with the rotational plane of the central compact hot core PA=140$^{o}$$\pm$5$^{o}$; Liu et al. 2010); and we make one PV cut through the UC H\textsc{ii} regions B, C, and the northern part of the biconical cavity (centered at \textbf{cutn: } R.A. = 18$^{h}$10$^{m}$27.45$^{s}$ , Decl = -19$^{o}$55$'$48.8$''$, with position angle 70$^{o}$).
Features we identify from these three PV cuts are described as follows.
Owing to the complexity of the observed region and the limited signal--to--noise ratio of the observations, there could be other dynamical features not revealed in the current maps.

\begin{figure}
\rotatebox{-90}{
\includegraphics[scale=0.55]{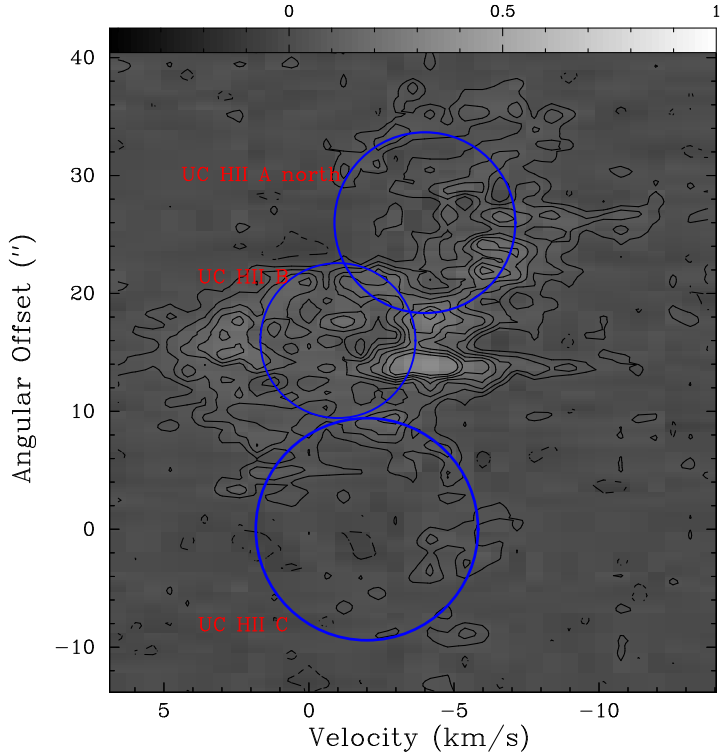}}
\caption{The PV diagram cuts through the UC H\textsc{ii} region B and C. The center of the PV cut is: R.A. = 18$^{h}$10$^{m}$28.51$^{s}$ , Decl = -19$^{o}$55$'$54.3$''$; the position angle of the cut is 70$^{o}$.  
Positive angular offset is defined in northeast.
Solid contours in both panels start from the brightness temperature of 11K  ($\sim$2$\sigma$), with 11K intervals; negative contours start from the brightness temperature of -11K, with -11K intervals. 
Blue circles from top to bottom are indicative of expansional signatures with centroid v$_{lsr}$=-4, -1 and -2 kms$^{-1}$, respectively; and the expansional velocity 3.1, 2.7, and 3.8 kms$^{-1}$, respectively.}
\label{fig_pv2}
\end{figure}

\paragraph{cutn} From the PV diagram of the cutn (Figure \ref{fig_pv2}), we identify the expansional signature in the northern part of the biconical cavity.
Around the UC H\textsc{ii} region B, we marginally see the arc--shaped signatures which have broad velocity width, and may also be explained as expansional signatures around the UC H\textsc{ii} region B.
The expansional velocities in these two regions are about 3 kms$^{-1}$. 
In these two regions, the expansional signatures are confused with the global rotation/contraction of the dense gas, and therefore the size and the expansional velocities are not robustly constrained.
The detection of the expansional signature in the UC H\textsc{ii} region C is marginal, with a suggested expansional velocity of 3.8 kms$^{-1}$.

\paragraph{cut1 and cut2}
Both the cut1 and cut2 are located south to the dense flattened structure; the center of cut2 is south of the center of cut1.
We argue that PV cuts in the position angle of 140$^{o}$ is less confused by the dynamics of the fast rotating ($\sim$3--4 kms$^{-1}$ rotational velocity) dense gas in the mid-plane of the envelope, and provides the clean information of the dynamics of the molecular cavity--wall.

\begin{figure}
\rotatebox{-90}{
\includegraphics[scale=0.9]{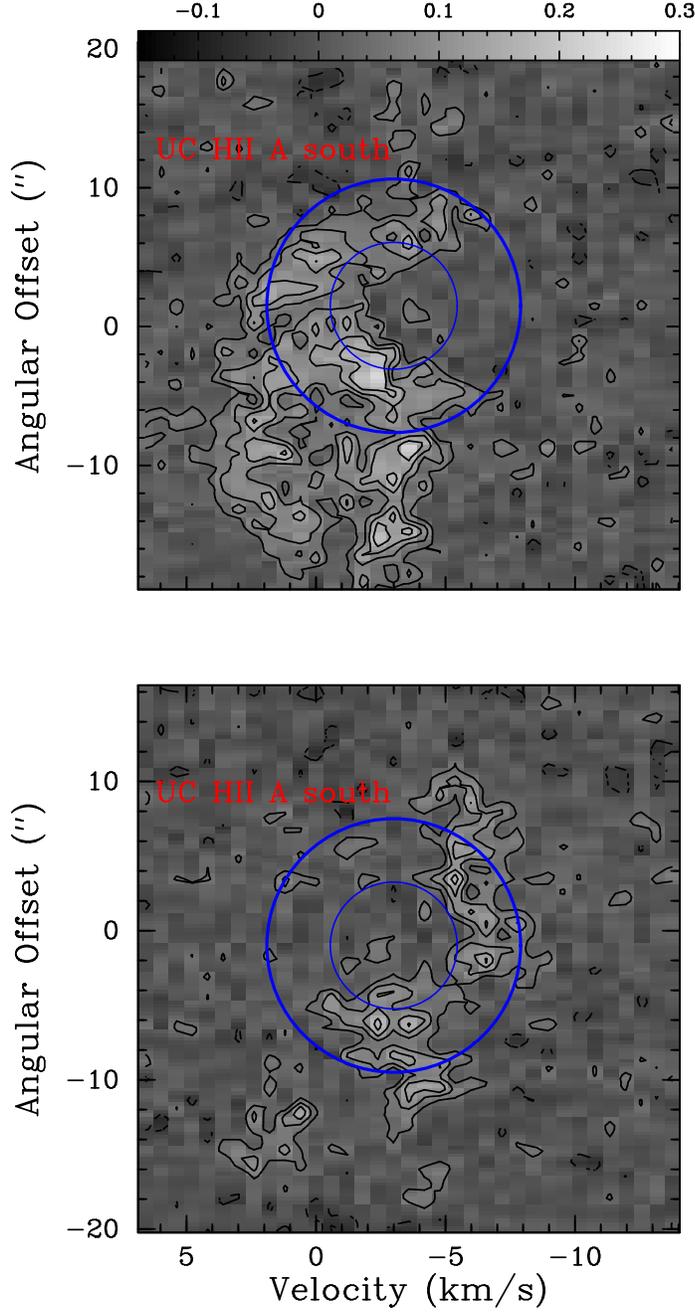}}
\caption{
\textbf{Top: }
The PV diagram cuts through the southwest lobe (0$''$ angular offset corresponds to R.A. = 18$^{h}$10$^{m}$28.51$^{s}$ , Decl = -19$^{o}$55$'$54.3$''$) with position angle 140$^{o}$. Positive angular offset is defined in southeast.
\textbf{Bottom: }
The PV diagram cuts through the southwest lobe (0$''$ angular offset corresponds to R.A. = 18$^{h}$10$^{m}$28.48$^{s}$ , Decl = -19$^{o}$56$'$1.0$''$) with position angle 140$^{o}$. 
Solid contours in both panels start from the brightness temperature of 11K, with 11K intervals; negative contours start from the brightness temperature of -11K, with -11K intervals. 
Blue circles are indicative of an expansional signature with centroid v$_{lsr}$=-3kms$^{-1}$, and the expansional velocity 2.5 and 5.0 kms$^{-1}$, respectively.
}
\label{fig_pv}
\end{figure}

The top and bottom panels of Figure \ref{fig_pv} show the PV diagram of cut1 and cut2, respectively.
The CS (1--0) emission in both panels clearly show arc--shaped expansional signatures centered at v$_{lsr}$$\sim$-3 kms$^{-1}$, with an averaged brightness temperature of 20--30 K, and the peak brightness temperature of 45 K.
Supposedly CS (1--0) is optically thick at the peak, then the brightness temperature represents the excitation temperature of the molecular gas.
The maximum excitation temperature of 45 K in the projected area of the biconical cavity explains why the CH$_{3}$OH transitions are difficult to be excited and have low optical depth. 
The projected linear scale of the expansional signature is about 10$''$ (0.3 pc), which is consistent with the scale of the visually identified biconical features in the CS (1-0) maps (Figures \ref{fig_cs}, \ref{fig_csx}), and the CH$_{3}$OH maps (Figures \ref{fig_nh3}, \ref{fig_ionout}). 

The arc--shaped signatures in the PV diagrams have a different physical size scale and geometry from the typical outflow signatures (Liu, Ho, \& Zhang 2010).
The arc--shaped feature shown by the cut1 may be explained as the redshifted  expansional motion of the cavity--wall.
Since the center of cut1 is closer to the free-free continuum peak, its blueshifted counter part may be affected by the free-free continuum subtraction, and therefore we cannot obtain significant detections; or, it can be explained by the non-uniformity of excitation temperature and density distribution (the 1$\sigma$ detection limit is 5.9 K).
The arc--shaped feature shown by cut2 can be explained as the blueshifted expansional motion.
The lack of the redshifted counter part of the expansional motion can either be explained by the asymmetry of molecular gas density, or asymmetrical stellar ionization. 
The line--of--sight expansional velocity of both features is about 2.5--5 kms$^{-1}$.
This interpretation can be constrained by observing the lower density tracers (e.g. $^{13}$CO (2-1) or C$^{18}$O (2-1) ), to sample the lower density/temperature part of the cavity wall.  
The emissions of those lower density tracers are extended, and the single dish observations will be needed in order to robustly interpret the data.

\section{Discussions}
\label{chap_discussion}

\subsection{The Overall Picture of the O--Type Cluster Forming Region}
\label{chap_overall}
\begin{figure}
\includegraphics[scale=0.75]{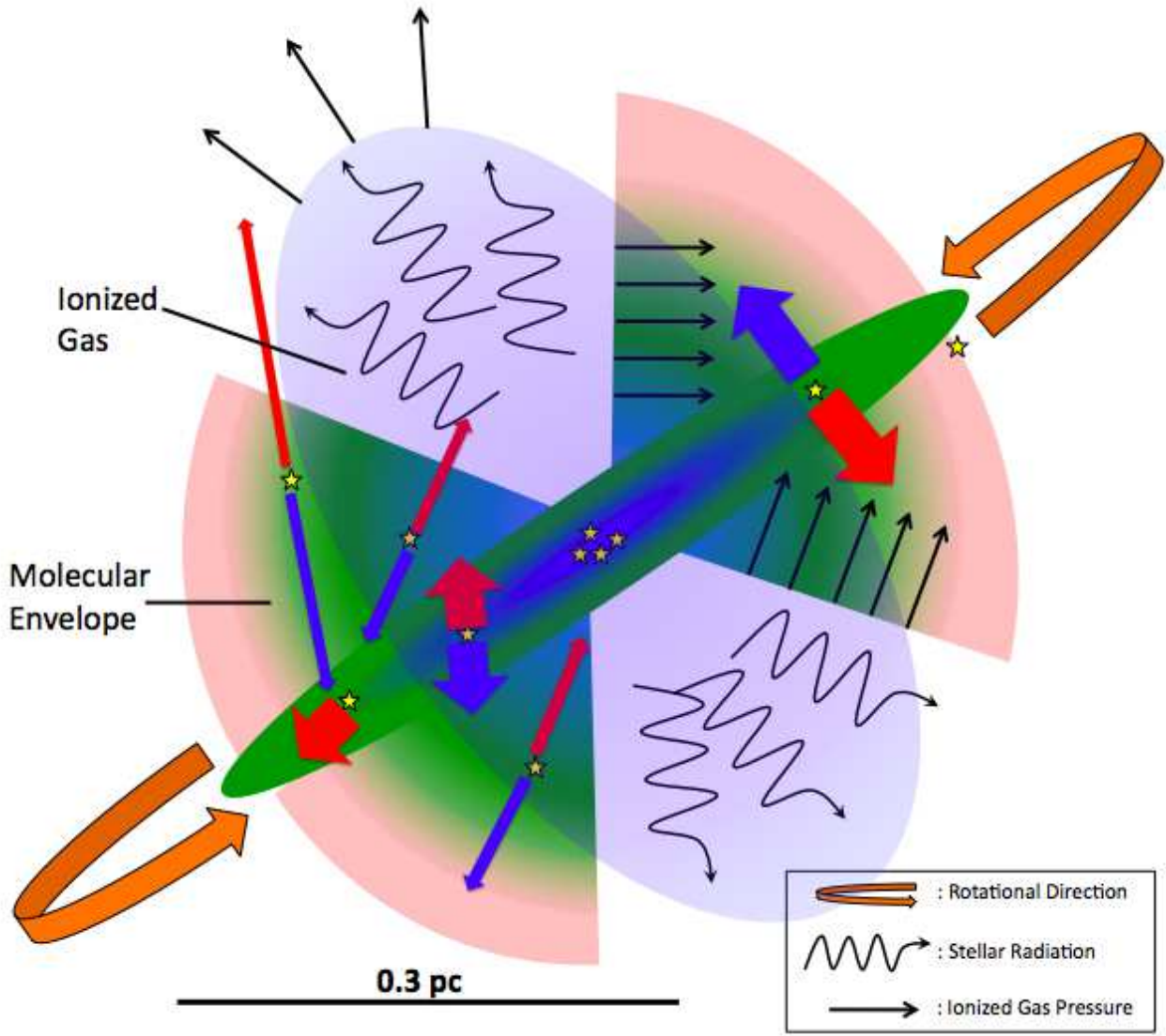}
\caption{A schematic view of the O--type cluster forming massive clump. The yellow star symbols indicates the (proto--)stellar objects. A compact O--type cluster is embedded in the center of the system, and is the dominant source of the stellar radiation. The Red and blue arrows indicate molecular outflows. Longer arrows represent outflows with larger terminal velocities;  the outflow momenta are proportional to the area of the arrows. We note that the molecular structures in this figure are highly clumpy/filamentary.}
\label{fig_shematic}
\end{figure}

We summarize our series of high resolution observations on G10.6--0.4 by a schematic model, which is shown in Figure \ref{fig_shematic}.
The observational results enable the discussions of the importance of the radiation and the pressure force of the ionized gas. 
In Sections \ref{chap_radiation} and \ref{chap_timescale}, we provide the order--of--magnitude estimates of the radiative pressure and the pressure forces of the ionized gas and the stellar wind.
We discuss their implications for the dynamical evolution of the system in Section \ref{chap_dp}.
We are looking forward to comparing our results with numerical hydrodynamical simulations in the future.

The dominant molecular structure in this model is a 0.5 pc scale massive envelope.
The massive envelope rotates and contracts globally.
A flattened overdensity formed in the mid--plane of the massive envelope (this is supported by the CS (1--0) observations, and also the $^{13}$CS (5--4) observations in Liu et al. 2010).
Meanwhile, the molecular gas also contracts locally, and forms intermediate--mass or B--type massive (proto--)stars, and some low--mass (proto--)stars.
The protostellar objects eject molecular outflows with high momentum and high energy into the ambient environment, which may affect the global contraction of the massive envelope (Li \& Nakamura 2006; Nakamura \& Li 2007; Carroll et al. 2009; Wang et al. 2010). 
Those molecular outflows can be revealed as highly blueshifted  and redshifted $^{12}$CO emissions ($|v|$$\ge$20 kms$^{-1}$) if the protostellar objects are not located in the mid--plane (Liu, Ho, \& Zhang 2010). 
In the mid--plane, the outflows are impeded and decelerated by the high density ambient gas to have lower terminal velocities, which have to be diagnosed from the molecular outflow/shock tracers (Liu, Ho, \& Zhang 2010).

The regulated global contraction may lead to the formation of the O--type cluster in the center (Li \& Nakamura 2006), which creates an UC H\textsc{ii} region.
The ionized gas and the ionizing photons leak out from the H\textsc{ii} region mainly in the bipolar direction, either because of a configuration where the rotationally flattened system has a low gas density in the bipolar region, or because of a strong protostellar MHD wind which created the biconical cavities in that region.
The outflowing molecular/ionized gas may flush through the ambient molecular gas on a timescale comparable to or much shorter than the global dynamical timescale\footnote{The lower limit of the global dynamical timescale can be estimated by the free--fall timescale of $\sim$10$^{5}$ years.}, and leave behind the elongated relics of interactions.  
This may explain the detected large filamentary structures by CS (1--0) in the northeast  (Section \ref{chap_csdis}).
Due to the short expansional timescale, the filamentary structures were not yet relaxed by the global rotational/infall motion. 
The pressure of the ionized gas can also push the cavity--wall to a quick dynamical expansion (Section \ref{chap_pv}).

However, in the rotationally flattened system, most of the molecular mass are concentrated in the mid--plane.
Owing to the self--gravitational or the Rayleigh--Taylor instabilities, the molecular gas becomes clumpy (Sollins \& Ho 2005), and additionally reduces the effective solid angle as presented to the central stars.
In such cases, the momentum impulse exerted by the ionized gas would be much smaller because of this smaller solid angle.
The inertia of the molecular accretion flow in the mid--plane can  easily overcome the impulse exerted by the stellar radiative pressure (also see Section \ref{chap_discussion}) and the pressure of the ionized gas.
This is consistent with the lack of a clear expansional motion in the PV diagram of the molecular lines  (Figure \ref{fig_ch3ohpv}).
When the molecular gas gets sufficiently close to the central OB cluster, it is heated by the stellar radiation to a significantly higher temperature, and is further ionized to a temperature of 10$^{4}$ K (also see discussions in Keto 2002; Keto \& Wood 2006).
This overall scenario allows the O--type stars to continue accreting after the nuclear burning begins.

\subsection{The Role of the Biconical Cavity in Enhancing the Molecular Accretion}
\label{chap_radiation}
The biconical cavity provides a low opacity channel for the photons to leak out of the envelope.
This mechanism can significantly reduce the radiative pressure on the molecular gas, and can therefore enhance the molecular accretion flow.

To quantitatively examine this effect, we first compare the relative importance of radiative pressure and the gravitational force in a system without the biconical cavity.
Following the derivations in Jijina \& Adams (1996), we assume the radiation field is nearly isotropic; and the stellar mass predominantly contributes to the gravitational force in the relevant scale.
We assume the temperature distribution follows the power law $T(r)=T_{0}(r/r_{0})^{-1/2}$ (Keto, Ho, \& Haschick 1987), and quote the resulting effective potential from Jijina \& Adams (1996):
\begin{equation}
V_{eff} = \frac{GM}{r}\left\{ \alpha r^{-1/2}-1\right\},
\label{eq1}
\end{equation}
where $M$ is the mass of the embedded stellar cluster.
The parameter $\alpha$ is defined by
\begin{equation}
\alpha\equiv\frac{L\kappa_{P}(T_{0})}{6\pi GMc}\sqrt{r_{0}},
\end{equation}
where $L$ is the bolometric luminosity of the embedded stellar cluster, $\kappa_{P}(T_{0})$ is the Planck mean opacity at the fiducial temperature $T_{0}$, and $r_{0}$ is the radius where $T=T_{0}$.
The first term in equation \ref{eq1} represents the radiative pressure, and the second term represents the gravitational force.
From equation \ref{eq1}, we see that when $\alpha^{2}/r\sim1$, the radiative pressure is comparably important with the gravity.
The radiative pressure becomes dominant at smaller radii.
In G10.6-0.4, the molecular hot toroid with temperature about 300 K is detected inward of the 0.05 pc radius\footnote{The averaged temperature might be overestimated. A lower value of 87 K is reported by Beltr{\'a}n et al. (2011) by observing optically thinner lines. We adopt the previous higher value to provide an upper limit of the radiation.}.
To match the measured molecular gas temperature, we adopt\footnote{We express the temperature power law with a parameter $T_{d}$, which is the dust sublimation temperature. If the temperature is higher than $T_{d}$, the dust is sublimated and the radiative pressure is reduced. The temperature power law of $T(r)=T_{0}(r/r_{0})^{-1/2}$ potentially overestimates the temperature at the 0.3 pc radius by a factor of 2 since the temperature decay faster with radius in regions with lower opacity, and has to be fitted with a more negative power law index. However, our discussions about the radiative pressure focus on the inner region which has high opacity.} $T_{0}=T_{d}\sim$ 2300 K ($\kappa(T_{0})$$\sim$30 cm$^{2}$g$^{-1}$) and $r_{0} =$ 8.5$\cdot$10$^{-4}$ pc. 
Given the G10.6-0.4 bolometric luminosity $L\sim$10$^{6}$ L$_{\odot}$ and stellar mass $M\sim$200 M$_{\odot}$, we obtain $\alpha^{2}\sim$0.07 pc (Ho \& Haschick 1981; Keto 2002; Sollins \& Ho 2005; Keto \& Wood 2006).
In such a case, the radiative pressure can significantly affect the dynamics at the scale of the hot toroid, or even reverse an inflow.

However, based on the numerical radiation transfer calculations of a set of systems with different geometries, Krumholz, McKee and Klein (2005) suggested that the presence of the biconical outflow cavity allows the photons to leak out, and therefore potentially reduces the radiative pressure force by a factor of $\sim$10.
This effectively reduce $L$ by a factor of 10, and therefore reduce $\alpha^2$ by a factor of 100.
Assuming this conclusion can be generally applied, with the presence of the biconical cavity,  the value of $\alpha^{2}$ in the case of G10.6--0.4 might be effectively reduced to 7$\cdot$10$^{-4}$ pc (140 AU), which is much smaller than the radius of the  UC H\textsc{ii} region in G10.6--0.4 ($\sim$0.03 pc). 
Qualitatively this estimate is consistent with the case in  Krumholz, McKee \& Klein (2005), suggesting that without considering the ionization, the radiative pressure will only be important on the scale of a few hundred AU.
We therefore suggest that in G10.6--0.4, the radiative pressure is unimportant for the dynamics of the molecular accretion flow.
We note that we may overestimate the reduction factor of the radiative pressure owing to the non-self-consistent treatment of the temperature distribution.
However, since $\alpha^{2}$ scales as the square of the reduction factor, it can easily become smaller than the radius of the H\textsc{ii} region with a small reduction factor. 
The presence of the biconical cavity can also lead to a temperature gradient  steeper than our assumption, and may makes $\alpha^{2}$ shrink to an even smaller radius.

The dusty disk around the stars may redirect the radiation (Yorke \& Sonnhalter 2002; Kuiper et al. 2010), and potentially play the role of reducing the radiative pressure on the accretion flow.
However, it is still uncertain whether the disks can stably exist in UC H\textsc{ii} regions, where multiple O--type stars are embedded. 
This has to be examined in future observations.

\subsection{The Ionized Gas Pressure and the Stellar Wind Feedback}
\label{chap_timescale}
The large molecular mass in the contracting massive envelope contains enormous inertia.
From the PV diagrams (Figure \ref{fig_ch3ohpv}), we find that the rotational velocity at the 0.25 pc radius is about 4--5 kms$^{-1}$.
This rotational motion is gravitationally bound by an enclosed mass of 1000--1600 M$_{\odot}$. 
While the embedded OB stars contribute $\sim$200 M$_{\odot}$, the rest of the binding mass is dominantly contributed by the molecular gas.
For a spherical geometry, this molecular mass corresponds to a mean molecular density $\bar{n_{H_{2}}}$ of (3--5)$\cdot$10$^{5}$ cm$^{-3}$.
The momentum flux in the molecular gas flow can be estimated by $\mu$$\cdot$$\bar{n_{H_{2}}}$$\cdot$$v^{2}$, where  $v$ is the infall velocity of the molecular gas, which can be larger than 1 kms$^{-1}$ (Ho \& Haschick 1986; Keto, Ho \& Haschick 1987, 1988; Keto 1990; Klaassen \& Wilson 2008),  and $\mu$ is the mean molecular weight.
In the embedded UC H\textsc{ii} region, the momentum flux in the ionized gas can be estimated by $\mu_{i}$$\cdot$$n_{i}$$\cdot$$v_{i}^{2}$, where $\mu_{i}$ is the ion mass, $n_{i}$ is the ion density, and $v_{i}^{2}$ is the thermal velocity of the ions, which has the order of magnitude of 10 kms$^{-1}$.
We assume that the ion density $n_{i}$ is equal to the electron density $n_{e}$, which is constrained to be 10$^{3}$--10$^{4}$ in previous observations (Ho and Haschick 1981; for a more sophisticated model fitting see Keto, Zhang, \& Kurtz 2008). 
Assuming $\mu$$\sim$2$\mu_{i}$, we find that the momentum flux in the molecular gas flow (6--10$\cdot$10$^{5}$$\cdot$$\mu_{i}$ M$_{\odot}$kms$^{-1}$cm$^{-2}$s$^{-1}$) and the momentum flux in the ionized gas (1--10$\cdot$10$^{5}$$\cdot$$\mu_{i}$ M$_{\odot}$kms$^{-1}$cm$^{-2}$s$^{-1}$)  have the same order of magnitudes.
This suggests that the two components are in an approximate dynamical equilibrium.
In the real case, the mass is anisotropically distributed. 
The dynamics of the accretion flow in the denser region will not be significantly affected by the pressure of the ionized gas, while the lower density region can undergo a dynamical expansion powered by the pressure of the ionized gas.

We estimate the momentum budget of the ionized gas and the stellar wind in this section.
In the next section, we will suggest a simple geometric model for the density distribution, and estimate the velocity of the dynamical expansion, to compare with the observations. 
The embedded massive stars in UC H\textsc{ii} regions B and C have similar spectral type (Ho \& Haschick 1981) with the embedded OB stars in UC H\textsc{ii} region A.
Some of the energetic relations estimated in UC H\textsc{ii} region A can also be applied in UC H\textsc{ii} region B and C.

\paragraph{Expansional Time Scale} 
The expansional signatures of the biconical cavity have a characteristic radius $r_{c}$ of about 5$''$. 
Assuming the fastest expanding front of the cavity--wall uniformly expands with the terminal velocity $v_{e}$ of 5 kms$^{-1}$, the total expanding time $t_{e}$ is about 3$\cdot$10$^{4}$ years, which is much shorter than the global dynamical timescale  ($>$10$^{5}$ years). 
The timescale estimation constrains the total momentum budget from the stellar wind and the ionized gas.

\paragraph{The Stellar Wind Feedback}
The UC H\textsc{ii} region A contains a few O6--O9 stars with a total stellar mass M$_{*}$ of $\sim$200 M$_{\odot}$.
The stellar wind from each of these massive stars has a mass loss rate of the order of magnitude of 10$^{-6}$ M$_{\odot}$yr$^{-1}$, and terminal velocity of $\sim$2000 kms$^{-1}$ (Tout et al. 1996).
Given the number of the embedded O stars $N_{*}$ ($\sim$4; Ho \& Haschick 1981), the wind momentum feedback rate of the massive cluster $f_{w}$ is $N_{*}$$\cdot$2$\cdot$10$^{-3}$ M$_{\odot}$kms$^{-1}$yr$^{-1}$.
Assuming the embedded stellar cluster starts to feed back the stellar wind by the beginning of the expansion of the biconical cavity, the total wind momentum budget $I_{w}$ is $N_{*}$$\cdot$60 M$_{\odot}$kms$^{-1}$ isotropically spread in the entire 4$\pi$ solid angle.
The exerted pressure force has a $r^{-2}$ radial dependence.

\paragraph{The Ionized Gas Pressure}
We adopt the characteristic radius $r$ of 0.15 pc, and adopt the electron density $n_{e}$ = 10$^{3}$--10$^{4}$ cm$^{-3}$ according to the measurements in Ho \& Haschick (1981), to estimate the thermal pressure force of the ionized gas.
Assuming the ion density equals to the electron density, and has the thermal velocity of $\sim$10 kms$^{-1}$, the momentum feedback rate $f_{i}$ can be estimated by $\Omega\cdot r^{2}\cdot(\rho\cdot v\cdot v)$, where $\Omega$ is the characteristic solid angle.
As an order of magnitude estimation, we adopt $\Omega$=4$\pi$,  which leads to $f_{i}$$\sim$6$\cdot$10$^{-4}$--6$\cdot$10$^{-3}$ M$_{\odot}$kms$^{-1}$yr$^{-1}$. 
The total ionized gas momentum feedback $I_{i}=\int_{0}^{t_{e}} f_{i}(r(t)) dt$ depends on the expansion history, and has the order of magnitude of 10$^{1}$--10$^{2}$ M$_{\odot}$kms$^{-1}$.

The ionized gas feedback differs from the stellar wind feedback in the sense that it is a pressure effect (see also Figure \ref{fig_shematic}), which does not exert  force in the radial direction, but in the direction perpendicular to the cavity--wall.
In addition, the exerted pressure force depends only on the temperature and density of the ion, which is determined by the ionization, the recombination, and the pressure balance.

\subsection{The Dynamical Process}
\label{chap_dp}
From the observations of the 1.3 mm continuum emission, and the measurements of the rotational velocity of the molecular gas, we constrained the molecular mass inward of the 0.15 pc radius around the UC H\textsc{ii} region A to have the order of magnitude of 400 M$_{\odot}$.
We assume a simple geometry in this region, and perform order--of--magnitude estimates of the energetic relation.

\subsubsection{The Geometrical Model}
Assume the molecular gas has an initial mass density distribution $\rho_{0}$ before the creation and the dynamical expansion of the biconical cavity.
Defining $\eta$ to be the angular separation from the plane of rotation, if the system is approximately axisymmetric, $\rho_{0}$ can be represented by $\rho_{0}\equiv\rho_{0}(r, \eta)$. 
If $\rho_{0}(r)$ can be represented as a polynomial of $\cos(\eta)$,  for a rotationally flattened system, we expect the leading order dependence to be $\cos(\eta)$.
This suggests that about 70\% of molecular mass (280 M$_{\odot}$) are concentrated in the region with $|\eta|<$45$^{o}$.

After the creation and the dynamical expansion of the biconical cavity, the molecular mass initially distributed in the region $|\eta|>$45$^{o}$ (120 M$_{\odot}$) are accumulated on the geometrically thin cavity--wall, which has an opening angle of 90$^{o}$.
The assumption of the mass accumulation on the cavity--wall is consistent with the fast expansion (see Section \ref{chap_timescale} for the estimate of timescale).
The molecular gas initially distributed in the region $|\eta|<$45$^{o}$ forms a dense flattened structure with density distribution $\rho$.
We argue that the opening angle of 90$^{o}$ is a reasonable value while comparing with the geometrical picture revealed by the CH$_{3}$OH images (Figure \ref{fig_nh3}, \ref{fig_ionout}).  
The cavity--wall has two components, W$_{\perp}$ and W$_{\parallel}$, of which the surface area are perpendicular and parallel to the radial direction, respectively. 
Given the opening angle, the ratio of the surface area of these two components is approximately 1; we assume the ratio of the mass accumulated on these two components is also 1. 

\subsubsection{The Dynamical Expansion of Cavity Wall}

\paragraph{W$_{\parallel}$} 
This component has a cone shape, with the total mass of $\sim$60 M$_{\odot}$. For it to expand at 2.5--5 kms$^{-1}$, it requires a momentum of 150--300 M$_{\odot}$kms$^{-1}$.
From the estimations in the previous section, we see that this required momentum has the same order of magnitude of the feedback   from the ionized gas pressure.
The gravitational force acts in the radial direction and does not retard the expansion in the $\hat{\theta}$ direction.
We expect the stellar wind to have weak effects on the dynamics of W$_{\parallel}$ since the wind is parallel to the cavity--wall. 

\paragraph{W$_{\perp}$} 
This part is directly driven by the stellar radiation and the stellar wind, which are competing with gravity in the radial direction. 
At the radius of 0.15 pc, the stellar wind and the ionized gas pressure can have comparable effect on W$_{\perp}$.
Assuming the mass is 60 M$_{\odot}$,  the gravitational force has the order of magnitude of 2$\cdot$10$^{-3}$ M$_{\odot}$kms$^{-1}$yr$^{-1}$, which is also comparable to the force of the stellar wind and the ionized gas.  

Observationally this part of cavity--wall is not as prominently detected as W$_{\parallel}$. 
Since the embedded stellar cluster in the UC H\textsc{ii} region A has a high ionizing photon emission rate ($S$ = 10$^{49}$ s$^{-1}$; Ho \& Haschick 1981), which is able to ionize on the order of 10$^{3}$ M$_{\odot}$ of gas in the biconical region.
Even if the recombination rate marginally balances the ionization rate, the stellar ionization can still explain why the W$_{\perp}$ is not clearly detected.
Some local high density clumps which have higher recombination rate may be self-shielded from the stellar ionization and have long lifetime. 
Those high density clumps occupy small solid angles, and are less accelerated by the stellar wind and the pressure of the ionized gas. 
Some of those clumps can still fall toward the OB cluster owing to the gravitational attraction. 
The clumpy 3.6 cm free-free continuum emissions in the UC H\textsc{ii} region A, especially, northeastern to the emission peak, may be explained by those clumps externally ionized by the central OB cluster.   

We note that we may overestimate the dynamical expansion timescale of the biconical cavity since the cavity can start expansion from a finite size owing to the initial ionization, which leads to the overestimation of the momentum feedback.
However, those ionized gases are not gravitationally bound and are free to leak out.
Therefore, the accumulated mass on the cavity--walls W$_{\parallel}$ and W$_{\perp}$ is reduced if the initial ionization is efficient, and requires smaller momentum feedback to reach the final state velocity.
Similar arguments are valid if the biconical cavity is initially created by the massive bipolar outflow (an example see G240.31+0.07: Qiu et al. 2009) and the molecular gas is evacuated.

\paragraph{The Dense Flattened Structure}
This structure has the mass of $\sim$280 M$_{\odot}$, and continues from the outer of $\sim$0.15 pc radius to the inner $\sim$0.03 pc (Liu et al. 2010).
The total gravitational force on this structure depends on the embedded mass.
A lower limit of 9.6$\cdot$10$^{-3}$ M$_{\odot}$kms$^{-1}$yr$^{-1}$ can be given assuming that the majority of mass is distributed at the 0.15 pc radius.
This lower limit is already on the same order of magnitude as the feedback from the stellar wind and the ionized gas pressure; the actual value of the gravitational force should be much larger. 
Depending on the effective solid angle it occupies, the stellar wind and the ionized gas pressure may have negligible effect on the radial motion of the flattened structure.
With the biconical cavity structure which leads to the photon leakage, the radiative pressure is also negligible.
We therefore suggest that the dynamics in this region is dominated by gravity and rotation.

\paragraph{UC HII region B and C}
The stellar wind and ionized gas pressure feedback in UC H\textsc{ii} region B and C have the same order of magnitude as those in the UC H\textsc{ii} region A.
However, these two regions have lower initial gas density, which leads to a smaller accumulated mass on the expansional shell.
The observed expansional velocities in these two regions are comparable to but smaller than the thermal velocity of the ionized gas, suggesting that the ionized gas pressure is an important driving source of the dynamical expansion.
From the estimations in the previous section, the stellar wind feedback can be comparably important, depending on the detailed geometry of these system.

\section{Summary}
\label{chap_summary}
We present high resolution observations of molecular lines and free-free continuum for the UC H\textsc{ii} region  G10.6--0.4.
The resolved projected distributions of the molecular gas and the ionized gas suggest an overall picture consisting of an extended ($\sim$0.5 pc) envelope, a single compact ($\sim$0.1 pc) hot rotating toroid, and a biconical molecular cavity filled with ionized gas.
This overall geometry resembles the standard envelope--disk and protostellar outflow model for the low--mass star forming region.
With the presence of the biconical cavity, we suggest that the radiative pressure can be significantly reduced.
In the plane of rotation, at the scale larger than 1$''$ (0.03 pc), we see that the rotational motion of the dense gas is not severely disturbed, which consistently suggests that the radiation is not yet important for the molecular accretion flow. 
The stellar radiation may increase and play a more important role as the stellar mass is increased. 

From the observations of CS (1--0), we suggest that the  biconical cavity around the UC H\textsc{ii} region A  is undergoing an expansional motion, with velocity of 2.5--5 kms$^{-1}$.
We perform simple order of magnitude estimates, and suggest that the feedback from the ionized gas pressure can account for the required momentum of this dynamical expansion.
The expansional signatures are also detected in the UC H\textsc{ii} region B and C, which are driven by the ionized gas pressure and the stellar wind.
The expansional motions of the UC H\textsc{ii} regions significantly disturb the local dynamics of the molecular gas, inject the energy, and may induce the non-uniformity in the molecular accretion flow, which is an important feedback mechanism in the massive molecular clump.

\acknowledgments
{\it Facilities:} \facility{SMA, VLA/EVLA}


\clearpage

\begin{table}
\scriptsize{
\begin{tabular}{lrccc}
Transition & Frequency (GHz) &   E$_{up}$/k (K) & Eins. A. (s$^{-1}$)  & Note\\\hline\hline
NH$_{3}$ (3,3) main & 23.870129 & 124.5  & 2.56$\cdot$10$^{-7}$  \\\hline


CS (1-0)                 & 48.990055 & 2.35    & 1.75$\cdot$10$^{-6}$  \\\hline

CH$_{3}$OH 5(0,5)-4(0,4) E  & 241.700219  &  47.68 & 6.04$\cdot$10$^{-5}$  \\\hline

CH$_{3}$OH 5(0,5)-4(0,4) A+ & 241.791431  &  34.65 & 6.05$\cdot$10$^{-5}$ & R \\\hline

CH$_{3}$OH 5(-2,4)-4(-2,3) E & 241.904152 &  60.38 & 5.09$\cdot$10$^{-5}$  & G\\
 
CH$_{3}$OH 5 (2,3)-4(2,2) E  & 241.904645 &  57.27 & 5.03$\cdot$10$^{-5}$  \\\hline

CH$_{3}$OH 5(-3,3)-4(-3,2) E & 241.852352 &  96.93 & 3.89$\cdot$10$^{-5}$  & B\\\hline

\end{tabular}
}
\caption{Table of the selected molecular transitions. The quantum number of the transitions are listed in the first column. Their frequencies and upper-level energy are listed in the second and the third column. The Einstein A--Coefficient of each transition is listed in the fourth column. The last column notes the corresponding color of the CH$_{3}$OH lines in the RGB images (Figure \ref{fig_nh3}, \ref{fig_ionout}).}
\label{table_molecule_list}
\end{table}

\begin{table}
\hspace{-2.4cm}
\scriptsize{
\begin{tabular}{lcclcc}
Transition                               &  Synthesized beam           &   RMS noise (Jy/beam)   & Instruments  & uv sampling range ($k\lambda$)   & Observed Date      \\\hline\hline
NH$_{3}$ (3,3) main                &   1$''$.8$\times$1$''$.2 &  0.008 & VLA/EVLA C--array                & 2.2--270  & 2009.07.27\\\hline


CS (1-0)                                &   1$''$.5$\times$1$''$.1 &  0.022 & VLA/EVLA DnC--array           & 4--245   & 2009.09.27 \\\hline

CH$_{3}$OH 5(0,5)-4(0,4) E    &   1$''$.5$\times$1$''$.3 &  0.06   & SMA compact+very extended  & 6--393  & 2009.06.10/2009.07.12\\\hline

CH$_{3}$OH 5(0,5)-4(0,4) A+  &   1$''$.5$\times$1$''$.3 &  0.06  & SMA compact+very extended   & 6--393  & 2009.06.10/2009.07.12\\\hline

CH$_{3}$OH 5(-2,4)-4(-2,3) E  &   1$''$.5$\times$1$''$.3 &  0.06  & SMA compact+very extended    & 6--393  & 2009.06.10/2009.07.12\\
 
CH$_{3}$OH 5 (2,3)-4(2,2) E   &   1$''$.5$\times$1$''$.3 &  0.06  & SMA compact+very extended    & 6--393 \\\hline

CH$_{3}$OH 5(-3,3)-4(-3,2) E  &   1$''$.5$\times$1$''$.3 &  0.06  & SMA compact+very extended    & 6--393  & 2009.06.10/2009.07.12\\\hline

\end{tabular}
}
\caption{Table of the instrumental parameters of the selected molecular transitions. The NH$_{3}$ data has the velocity resolution of 1.2 kms$^{-1}$; the CS (1--0) data has the velocity resolution of 0.6 kms$^{-1}$; and the CH$_{3}$OH data has the velocity resolution of 0.5 kms$^{-1}$. The rms noises shown in the third column are measured from the channel images with corresponding velocity widths.}
\label{table_molecule_list2}
\end{table}

\begin{table}
\scriptsize{
\begin{tabular}{lccccc}
Observation                                        & Flux Cal.        &      Passband Cal   &       Gain Cal                    &  Total Bandwidth        &   Frequency Resolution   \\\hline\hline 
VLA/EVLA C--array (NH$_{3}$)             & 1331+105    &      3c273             &    1820-254                    & 6.25 MHz                   &    97.656 kHz                \\

VLA/EVLA DnC--array (CS)                 & 1331+105    &      0319+415       &    1733-130/1820-254    & 6.25 MHz                   &    97.656 kHz                \\

SMA compact--array                            & Uranus           &     3c273              &    1733-130/1911-201    & 2 GHz in each sideband &   406 kHz                      \\

SMA very--extended--array                   & mwc349        &     3c273              &    1733-130/1911-201    & 2 GHz in each sideband &   406 kHz                      \\\hline

\end{tabular}
}
\caption{The observational settings of the selected molecular transitions.}
\label{table_molecule_list3}
\end{table}

\clearpage

\clearpage

\begin{thebibliography}{}




\bibitem[Beltr{\'a}n et al.(2011)]{2011A&A...525A.151B} Beltr{\'a}n, M.~T., Cesaroni, R., Neri, R., \& Codella, C.\ 2011, \aap, 525, A151

\bibitem[Beuther et al.(2004)]{2004ApJ...616L..23B} Beuther, H., et al.\ 2004, \apjl, 616, L23



\bibitem[Carroll et al.(2009)]{2009ApJ...695.1376C} Carroll, J.~J., Frank, A., Blackman, E.~G., Cunningham, A.~J., \& Quillen, A.~C.\ 2009, \apj, 695, 1376

\bibitem[Caswell et al.(1975)]{1975A&A....45..239C} Caswell, J.~L., Murray, J.~D., Roger, R.~S., Cole, D.~J., \& Cooke, D.~J.\ 1975, \aap, 45, 239 

\bibitem[Corbel \& Eikenberry(2004)]{2004A&A...419..191C} Corbel, S., \& Eikenberry, S.~S.\ 2004, \aap, 419, 191

\bibitem[Downes et al.(1980)]{1980A&AS...40..379D} Downes, D., Wilson, T.~L., Bieging, J., \& Wink, J.\ 1980, \aaps, 40, 379

\bibitem[Frank, Balick \& Davidson(1995)]{1995ApJ...441L..77F} Frank, A., Balick, B., \& Davidson, K.\ 1995, \apjl, 441, L77

\bibitem[Frank, Ryu \& Davidson(1998)]{1998ApJ...500..291F} Frank, A., Ryu, D., \& Davidson, K.\ 1998, \apj, 500, 291



\bibitem[Galv{\'a}n-Madrid et al.(2008)]{2008ApJ...674L..33G} Galv{\'a}n-Madrid, R., Rodr{\'{\i}}guez, L.~F., Ho, P.~T.~P., \& Keto, E.\ 2008, \apjl,674, L33

\bibitem[Galv{\'a}n-Madrid et al.(2009)]{2009ApJ...706.1036G} Galv{\'a}n-Madrid, R., Keto, E., Zhang, Q., Kurtz, S., Rodr{\'{\i}}guez, L.~F., \& Ho, P.~T.~P.\ 2009, \apj, 706, 1036

\bibitem[Garay \& Lizano(1999)]{1999PASP..111.1049G} Garay, G., \& Lizano, S.\ 1999, \pasp, 111, 1049

\bibitem[Genzel \& Downes(1977)]{1977A&AS...30..145G} Genzel, R., \& Downes, D.\ 1977, \aaps, 30, 145

\bibitem[Guilloteau et al.(1988)]{1988A&A...202..189G} Guilloteau, S., Forveille, T., Baudry, A., Despois, D., \& Goss, W.~M.\ 1988, \aap, 202, 189   


\bibitem[Ho \& Haschick(1981)]{1981ApJ...248..622H} Ho, P.~T.~P., \& Haschick, A.~D.\ 1981, \apj, 248, 622

\bibitem[Ho et al.(1983)]{1983ApJ...265..295H} Ho, P.~T.~P., Vogel, S.~N., Wright, M.~C.~H., \& Haschick, A.~D.\ 1983, \apj, 265, 295 

\bibitem[Ho \& Townes(1983)]{1983ARA&A..21..239H} Ho, P.~T.~P., \& Townes, C.~H.\ 1983, \araa, 21, 239 

\bibitem[Ho \& Haschick(1986)]{1986ApJ...304..501H} Ho, P.~T.~P., \& Haschick, A.~D.\ 1986, \apj, 304, 501

\bibitem[Ho et al.(1986)]{1986ApJ...305..714H} Ho, P.~T.~P., Klein, R.~I., \& Haschick, A.~D.\ 1986, \apj, 305, 714 

\bibitem[Ho et al.(1994)]{1994ApJ...423..320H} Ho, P.~T.~P., Terebey, S., \& Turner, J.~L.\ 1994, \apj, 423, 320 
       
\bibitem[Ho et al.(2004)]{2004ApJ...616L...1H} Ho, P.~T.~P., Moran, J.~M., \& Lo, K.~Y.\ 2004, \apjl, 616, L1       

\bibitem[Hofner \& Churchwell(1996)]{1996A&AS..120..283H} Hofner, P., \& Churchwell, E.\ 1996, \aaps, 120, 283

\bibitem[Jijina \& Adams(1996)]{1996ApJ...462..874J} Jijina, J., \& Adams, F.~C.\ 1996, \apj, 462, 874 

\bibitem[Kahn(1974)]{1974A&A....37..149K} Kahn, F.~D.\ 1974, \aap, 37, 149

\bibitem[Keto et al.(1987)]{1987ApJ...318..712K} Keto, E.~R., Ho, P.~T.~P., \& Haschick, A.~D.\ 1987, \apj, 318, 712 

\bibitem[Keto et al.(1988)]{1988ApJ...324..920K} Keto, E.~R., Ho, P.~T.~P., \& Haschick, A.~D.\ 1988, \apj, 324, 920 
 
\bibitem[Keto(1990)]{1990ApJ...355..190K} Keto, E.~R.\ 1990, \apj, 355, 190 
 

\bibitem[Keto(2002)]{2002ApJ...568..754K} Keto, E.\ 2002a, \apj, 568, 754

\bibitem[Keto(2002)]{2002ApJ...580..980K} Keto, E.\ 2002b, \apj, 580, 980

\bibitem[Keto(2003)]{2003ApJ...599.1196K} Keto, E.\ 2003, \apj, 599, 1196

\bibitem[Keto \& Wood(2006)]{2006ApJ...637..850K} Keto, E., \& Wood, K.\ 2006, \apj, 637, 850

\bibitem[Keto(2007)]{2007ApJ...666..976K} Keto, E.\ 2007, \apj, 666, 976

\bibitem[Keto et al.(2008)]{2008ApJ...672..423K} Keto, E., Zhang, Q., \& Kurtz, S.\ 2008, \apj, 672, 423  

\bibitem[Keto \& Zhang(2010)]{2010MNRAS.406..102K} Keto, E., \& Zhang, Q.\ 2010, \mnras, 406, 102
 
\bibitem[Klaassen \& Wilson(2008)]{2008ApJ...684.1273K} Klaassen, P.~D., \& Wilson, C.~D.\ 2008, \apj, 684, 1273
 
\bibitem[Klaassen et al.(2009)]{2009ApJ...703.1308K} Klaassen, P.~D., Wilson, C.~D., Keto, E.~R., \& Zhang, Q.\ 2009, \apj, 703, 1308

\bibitem[Krumholz et al.(2005)]{2005ApJ...618L..33K} Krumholz, M.~R., McKee, C.~F., \& Klein, R.~I.\ 2005, \apjl, 618, L33

\bibitem[Krumholz et al.(2009)]{2009Sci...323..754K} Krumholz, M.~R., Klein, R.~I., McKee, C.~F., Offner, S.~S.~R., \& Cunningham, A.~J.\ 2009, Science, 323, 754 

\bibitem[Kuiper et al.(2010)]{2010ApJ...722.1556K} Kuiper, R., Klahr, H., Beuther, H., \& Henning, T.\ 2010, \apj, 722, 1556

\bibitem[Leurini et al.(2010)]{2010A&A...511A..82L} Leurini, S., Parise, B., Schilke, P., Pety, J., \& Rolffs, R.\ 2010, \aap, 511, A82

\bibitem[Li \& Nakamura(2006)]{2006ApJ...640L.187L} Li, Z.-Y., \& Nakamura, F.\ 2006, \apjl, 640, L187


\bibitem[Liu et al.(2010)]{2010ApJ...722..262L} Liu, H. B., Ho,  P.~T.~P., Zhang, Q., Keto, E., Wu, J., \& Li, H.\ 2010, \apj, 722, 262

\bibitem[Liu et al.(2010)]{2010ApJ...725.2190L} Liu, H.~B., Ho, P.~T.~P., \& Zhang, Q.\ 2010, \apj, 725, 2190


\bibitem[Nakamura \& Li(2007)]{2007ApJ...662..395N} Nakamura, F., \& Li, Z.-Y.\ 2007, \apj, 662, 395


\bibitem[Omodaka et al.(1992)]{1992PASJ...44..447O} Omodaka, T., Kobayashi, H., Kitamura, Y., Nakano, M., \& Ishiguro, M.\ 1992, \pasj, 44, 447    

\bibitem[Peters et al.(2010)]{2010ApJ...711.1017P} Peters, T., Banerjee, R., Klessen, R.~S., Mac Low, M.-M., Galv{\'a}n-Madrid, R., \& Keto, E.~R.\ 2010, \apj, 711, 1017



\bibitem[Shepherd et al.(2001)]{2001Sci...292.1513S} Shepherd, D.~S., Claussen, M.~J., \& Kurtz, S.~E.\ 2001, Science, 292, 1513

\bibitem[Sollins et al.(2005)]{2005ApJ...624L..49S} Sollins, P.~K., Zhang, Q., Keto, E., \& Ho, P.~T.~P.\ 2005, \apjl, 624, L49           

\bibitem[Sollins \& Ho(2005)]{2005ApJ...630..987S} Sollins, P.~K., \& Ho, P.~T.~P.\ 2005, \apj, 630, 987   


\bibitem[Takakuwa et al.(2003)]{2003ApJ...590..932T} Takakuwa, S., Ohashi,  N., \& Hirano, N.\ 2003, \apj, 590, 932

\bibitem[Wang et al.(2010)]{2010ApJ...709...27W} Wang, P., Li, Z.-Y., Abel, T., \& Nakamura, F.\ 2010, \apj, 709, 27


\bibitem[Yorke \& Sonnhalter(2002)]{2002ApJ...569..846Y} Yorke, H.~W., \& Sonnhalter, C.\ 2002, \apj, 569, 846




\bibitem[Zhang et al.(1998)]{1998ApJ...505L.151Z} Zhang, Q., Hunter, T.~R., \& Sridharan, T.~K.\ 1998, \apjl, 505, L151

\end{thebibliography}
\end{document}